\documentclass[preprint,12pt]{elsarticle}

\usepackage{amssymb}
\usepackage{url}
\usepackage{amsmath}
\usepackage{enumerate}
\usepackage{bm}
\usepackage{fancyhdr}
\DeclareMathOperator\erf{erf}

\makeatletter
\def\ps@pprintTitle{%
 \let\@oddhead\@empty
  \def\@oddhead{\copyright 2017. This manuscript version is made available under the CC-BY-NC-ND 4.0 license 

}%

 \def\@oddfoot{Preprint Accepted for Publication in Nuclear Instruments and Methods in Physics Research A}%
 \let\@evenfoot\@oddfoot}

\usepackage[nodots,nocompress]{numcompress}

\journal{Nuclear Instruments and Methods in Physics Research A }

\begin{document}

\begin{frontmatter}

\title{Instrument Performance and Simulation Verification of the POLAR Detector}
\author[dpnc]{M.~KOLE\corref{cor1}}
\ead{Merlin.Kole@unige.ch}
\author[ihep,cast]{Z.H.~LI}
\author[isdc]{N.~PRODUIT}
\author[pol]{T.~TYMIENIECKA}
\author[ihep]{J.~ZHANG}
\author[pol]{A.~ZWOLINSKA}

\author[ihep]{T.W.~BAO}
\author[isdc]{T.~BERNASCONI}
\author[dpnc]{F.~CADOUX}
\author[ihep,cast] {M.Z.~FENG}
\author[isdc]{N.~GAUVIN}
\author[psi]{W.~HAJDAS}
\author[ihep]{S.W.~KONG}
\author[ihep,cast]{H.C.~LI}
\author[ihep]{L.~LI}
\author[ihep]{X.~LIU}
\author[psi]{R.~MARCINKOWSKI}
\author[dpnc]{S.~ORSI}
\author[dpnc]{M.~POHL}
\author[pol]{D.~RYBKA}
\author[ihep]{J.C.~SUN}
\author[ihep]{L.M.~SONG}
\author[pol]{J.~SZABELSKI}
\author[ihep]{R.J.~WANG}
\author[ihep,cast]{Y.H.~WANG}
\author[ihep,cast]{X.~WEN}
\author[ihep]{B.B.~WU}
\author[dpnc]{X.~WU}
\author[psi]{H.L.~XIAO}
\author[ihep]{S.L.XIONG}
\author[ihep]{L.~ZHANG}
\author[ihep]{L.Y.~ZHANG}
\author[ihep]{S.N.~ZHANG}
\author[ihep]{X.F.~ZHANG}
\author[ihep]{Y.J.~ZHANG}
\author[ihep,lan]{Y.~ZHAO}

\address[dpnc]{University of Geneva (DPNC), quai Ernest-Ansermet 24, 1205 Geneva, Switzerland}
\address[ihep]{Key Laboratory of Particle Astrophysics, Institute of High Energy Physics, Chinese Academy of Sciences, Beijing, China, 100049}
\address[cast]{University of Chinese Academy of Sciences, Beijing 100049, China}
\address[isdc]{University of Geneva, Geneva Observatory, ISDC,
16, Chemin d'Ecogia, 1290 Versoix Switzerland}
\address[pol]{National Centre for Nuclear Research
ul. A. Soltana 7, 05-400 Otwock, Swierk, Poland}
\address[psi]{Paul Scherrer Institut
5232 Villigen PSI, Switzerland}
\address[lan]{School of Nuclear Science and Technology, Lanzhou University, Lanzhou 730000, China}

\cortext[cor1]{Corresponding author}

\begin{abstract}
POLAR is a new satellite-born detector aiming to measure the polarization of an unprecedented number of Gamma-Ray Bursts in the 50-500 keV energy range. The instrument, launched on-board the Tiangong-2 Chinese Space lab on the 15th of September 2016, is designed to measure the polarization of the hard X-ray flux by measuring the distribution of the azimuthal scattering angles of the incoming photons. A detailed understanding of the polarimeter and specifically of the systematic effects induced by the instrument's non-uniformity are required for this purpose. In order to study the instrument's response to polarization, POLAR underwent a beam test at the European Synchrotron Radiation Facility in France. In this paper both the beam test and the instrument performance will be described. This is followed by an overview of the Monte Carlo simulation tools developed for the instrument. Finally a comparison of the measured and simulated instrument performance will be provided and the instrument response to polarization will be presented.
\end{abstract}

\end{frontmatter}


\section{Introduction}
\label{sec:1}

POLAR \cite{Nicolas} is a dedicated Gamma-Ray Burst (GRB) polarimeter launched on September 15th 2016 on-board the Chinese Space lab Tiangong-2 (TG-2) into low earth orbit. It orbits the earth at an altitude of $400\,\mathrm{km}$ with an orbital inclination of $42.8^\circ$. The design of the flight model is described in detail in \cite{Nicolas_new}. The goal of the mission is to measure the polarization of a large number of GRBs in the 50-500 keV energy range. For this purpose the instrument's design incorporates a large field of view, a relatively large effective area and a high sensitivity to polarization. The polarization of the incoming X-ray flux is determined by measuring the distribution of the azimuthal Compton scattering angles. The Compton scattering angle of an X-ray is preferentially perpendicular to its polarization angle as described by the Klein-Nishina formula \cite{KNF}, presented in \ref{KN}:
\begin{equation} \label{KN}
\mathrm{\frac{d\sigma}{d\Omega}=\frac{r_0^2}{2}\left( \frac{E'}{E} \right)^2\left(\frac{E'}{E}+\frac{E}{E'}-2\sin^2\theta \cos^2\phi\right).}
\end{equation}

Here $\mathrm{r_0^2}$ is the classical electron radius, $\mathrm{E}$ is the initial photon energy, $\mathrm{E'}$ the final photon energy, $\theta$ the polar scattering angle and $\phi$ the azimuthal scattering angle with respect to the polarization vector. The distribution of the azimuthal Compton scattering angle of a polarized photon flux will be a harmonic function with a $180^\circ$ period called a modulation curve which can be parameterized as:

\begin{equation} 
f (\eta) = T (1 + \mu (\cos(2\phi + 2\alpha)))
\end{equation}

Here T is the mean of a harmonic function with a 180 degree period, the parameter $\phi$ is the azimuthal angle and $\alpha$ is the polarization angle. The polarization degree $\Pi$ can be acquired from this function using: 

\begin{equation}
 \Pi=\frac{\mu}{\mu_{100}}.
\end{equation}

Here $\mu$ is the measured amplitude while $\mu_{100}$ is the amplitude as measured for a $100\%$ polarized beam. The value of $\mu_{100}$ is detector specific and can be acquired through dedicated measurements verified with Monte Carlo simulations. 

\begin{figure}[h!]
  \centering
    \includegraphics[width=12 cm]{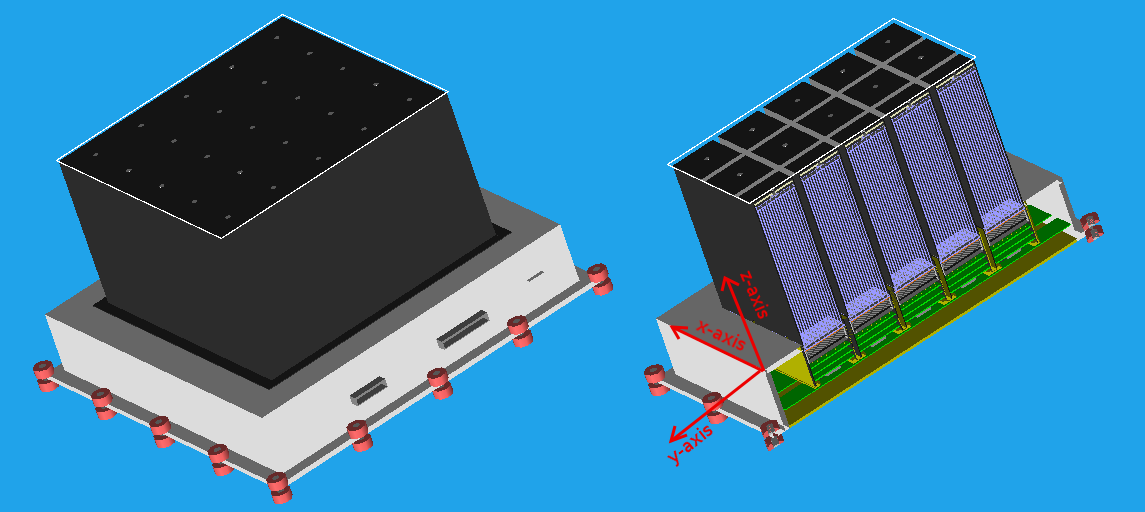}
  \caption{A schematic representation of the POLAR detector. On the left the full instrument is displayed. On the right the instrument cut in two along the y-z plane is shown and the carbon cover is removed to make the individual DMs visible. The scintillator bars are visible in the right figure. The bars are surrounded on the top and side by a carbon fiber shield and on the bottom by the PMTs (dynode planes can be seen) and the electronics. The carbon shielding is connected to an aluminum frame which is connected to the TG-2 through the ring shaped rubber dampers (red).}
\label{fig:SIM}
\end{figure}

In POLAR the Compton scattering angle is measured using a segmented scintillator array consisting of a total of 1600 plastic scintillator bars. These scintillators, of the type EJ-248M \cite{EJ248M}, have a length of $176\mathrm{mm}$ and a square cross section of $5.8\times5.8\,\mathrm{mm}$. The scintillators are read out in groups of 64 using an $8\times8$ Multi-Anode PMT (MAPMT) from Hamamatsu. Each MAPMT is read out using its own front-end electronics (FEE), which together with the MAPMT and the scintillator, forms a detector module (DM). The combination of a scintillator bar and its corresponding readout channel will from here on be referred to as a bar. The full detector consists of an array of $5\times5$ DMs. A schematic representation of the full instrument and a cross section of the the POLAR detector can be seen in figure \ref{fig:SIM}.

As the polarimeter is designed to measure the Compton scattering angle, an event in the detector typically consists of a Compton scattering interaction in one scintillator followed by a second Compton scattering interaction or photo-absorption of the photon in a second scintillator. As there is no position resolution within the scintillator bars the scattering angle is measured as the angle between two random positions, one in each bar. A single FEE is triggered in case one of its 64 readout channels exceeds a set threshold. In this case the FEE sends a signal containing the number of triggered bars to the central trigger (CT) system which takes the decision on the readout and, if valid, sends the readout command to all DMs with a trigger. The CT system reads out an event in case at least two bars are triggered in the full instrument. These two triggers can either be inside a single DM or in two separate DMs. The coincidence window for triggers in different bars is approximately $50\,\mathrm{ns}$. The dead time of a DM after an accepted trigger is $68\,\mathrm{\mu s}$, all other DMs can still accept events during this period \cite{Nicolas_new}. In order to limit the data size per event only DMs containing at least one triggering bar are read out. To reject events induced by charged cosmic rays the FEE issues a veto in case too many bars within a DM are above trigger level. An analogue signal proportional to the number of triggered bars within a DM is compared to an adjustable veto threshold. By default this threshold is equal to approximately 4 bars. In order to be able to study the veto efficiency a fraction of the vetoed events is stored, these events are referred to as prescaled cosmic events. During the beam tests the cosmic-ray rate was of the order of several 10's of Hz and therefore negligible. In orbit this rate is expected to be an order of magnitude larger \cite{Estela}. The same logic is applied to events which consist of a single trigger in the full instrument. These events are referred to as prescaled single events. 

As POLAR is designed to measure the polarization of emission from transient events the instrument has a large field of view which covers about half the sky. An X-ray flux entering the instrument along the z-axis, so in the length direction of the scintillators, produces a relatively undisturbed modulation curve. However the modulation curve, with a period of $180^\circ$, will be affected by geometrical and instrumental effects. Instrumental effects are for example variations in gain and trigger thresholds for different bars. A first geometrical effect is due to the majority of the GRBs appearing off-axis. This will result in a harmonic function with a $360^\circ$ period on top of the modulation induced by polarization. Additionally the square geometry of the scintillator bars and the lack of position resolution within the scintillator bar results in a larger likelihood to measure scattering aligned along the x or y-axis. As a result a 3rd harmonic function with a $90^\circ$ period is superimposed on top of the $180^\circ$ and $360^\circ$ degree harmonic functions. Such geometrical effects were found in previous beam tests performed on a single DM of POLAR presented in \cite{Silvio}. In order to study these geometrically and instrumentally induced effects on the modulation curves for the full flight model of POLAR, the detector underwent a photon beam test at the European Synchrotron Radiation Facility (ESRF) in Grenoble, France, in May 2015. These tests and the configuration are described in the next section. This is followed by an overview of the Monte Carlo simulations used to simulate the instrument. Finally the beam test results are used to verify the simulations such that these can be used to correct for the geometrical effects in GRB induced modulation curves.

\section{Beam Test}

\begin{figure}[h!]
  \centering
    \includegraphics[width=9 cm]{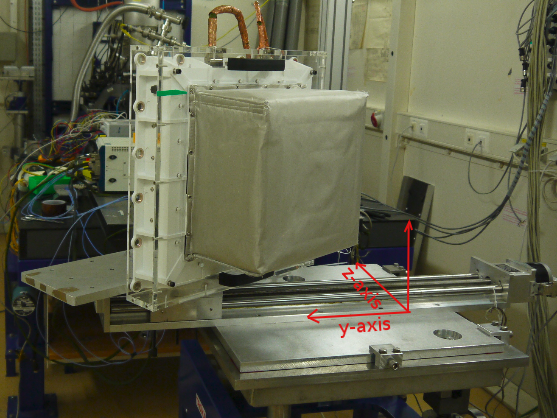}
  \caption{The POLAR detector in the ID11 beam hutch at ESRF during a 30 degree off-axis irradiation. The beam is directed along the z-axis, while the instrument is moved automatically during a scan along the x and y directions. The polarization of the beam is along the y-axis.}
\label{fig:table}
\end{figure}

In May 2015 beam tests were performed with the flight model of the POLAR detector at ESRF in Grenoble, France. In order to mimic a uniform irradiation of the instrument as close as possible using a relatively small beam the detector was placed on a motorized table designed to move the instrument during irradiation. This setup can be seen in figure \ref{fig:table}. The table was setup for the beam to move through the center of each bar once during a scan. An equal irradiation of each bar was ensured by scanning one row of bars in the y-direction, as defined in the figure. Subsequently while the instrument was not directly illuminated the detector was moved up in the x-direction by a distance of $6.08\,\mathrm{mm}$, corresponding to the center to center distance between two bars. A full instrument scan therefore consists of the combination of 40 scans in the y-direction, one of each row of bars. For tests where the instrument was illuminated on-axis the motor moving the table along the y-direction was set to a velocity of $0.5\,\mathrm{mm/s}$ corresponding to an illumination of each bar of $1.16\pm0.01\,\mathrm{s}$. The beam size was set to be its maximum which corresponds to $0.6\times0.6\,\mathrm{mm}$, well within the cross section of a bar. Careful alignment of the beam was performed at the start of each run to ensure the beam to move as close as possible through the center of each bar. The uncertainty on the alignment is of the order of 1 mm. The trigger rate measured by the instrument during the beam test was similar to that measured in-orbit during a GRB.

The detector was irradiated with 4 different beam energies, $60, 80, 110$ and $140\,\mathrm{keV}$ at the ID11 beam line at ESRF. The bandpass, $\mathrm{dE/E}$, of the monochromator used at this beamline is $2\times10^{-3}$ \cite{ESRF}. Due to some issues with the scans with $140\,\mathrm{keV}$ only the first 3 energies will be considered in this paper. For each energy 4 different scans were performed. Two on-axis scans, where the detector was rotated between scans by $90$ degrees around the z-axis, were performed. These measurements allow both for measurements of the polarization in two orthogonal directions and for a measurement of a non-polarized beam by combining the data from the two scans. Subsequently two scans were performed where the detector was rotated by $30$ and $60$ degrees around the x-axis, allowing for testing the instrument performance to off-axis beams. 

For each set of measurements the beam energy and position was calibrated using a germanium detector provided by ESRF which could be moved into the beam. This detector was also used to verify that no significant background, e.g. from scattering of dead materials in the hutch, was present during the instrument irradiation. The lack of background induced by the beam was also verified by studying data from periods during which the beam was not irradiating the instrument while the beam shutter was moved into position to block the beam. No significant decrease exceeding the statistical fluctuations of the trigger rate was found while the shutter was moved into position. Indicating that no additional background is induced by the beam.

The synchrotron used to produce the beam for the calibration of POLAR was refilled several times per day. During refilling of the synchrotron the beam intensity increases by approximately $50\%$ within several seconds after which it gradually decreases again until the next refilling. As a result the beam intensity and therefore the current as measured in the synchrotron, will vary with time. The beam current, which is proportional to the photon beam flux, was stored once per second during the full calibration campaign. Along with the beam current the time and the positions of the x and y translators used to move the detector were also stored once per second. This allows for the application of a correction for the beam current fluctuations in the analysis phase.

During the beam test POLAR was operated in the same mode as in-orbit. Data storage and communication with the instrument was performed using copies of the instrumentation currently used in-orbit for this purpose. It should however be noted that during the beam test the instrument suffered from several minor firmware (FW) related issues which have since been resolved. One relevant firmware related issue encountered during the tests resulted in sporadic bad initialization of the time stamps recorded on the FEE of modules. As a result coincident triggers for events recorded in such a DM and other DMs in the instrument were therefore lost. DMs suffering from this issue will therefore not be considered in the analysis. Furthermore 5 of the 25 DMs used during the calibration have been replaced before launch as these DMs suffered from other types FEE related issues. For two of these DMs the problems were already present during the beam test and these will therefore additionally not be considered in the analysis. 

\section{Data Analysis}

Two significant differences between the ESRF data analysis and the foreseen flight data analysis consist of the energy calibration procedure which is discussed in the following subsection and the correction for the variations in the beam current. Furthermore as described in the previous section several DMs were excluded from analysis due to problems with the FEE or with the FW during the specific run. In-orbit no such issues with the DMs are present.

\subsection{Energy Calibration}

During in-orbit data taking the energy calibration of each bar is performed using 4 internal $^{22}\mathrm{Na}$ sources placed inside the instrument. As a positron emitter $^{22}\mathrm{Na}$ emits two back-to-back photons each with an energy of $511\,\mathrm{keV}$. The back-to-back nature of these photons together with the knowledge of their exact position in the polarimeter can be used to accurately select events from the sources. Using this event selection a $^{22}\mathrm{Na}$ induced spectrum can be produced for each bar. The Compton Edge equal to $341\,\mathrm{keV}$ is fitted and used for energy calibration. With a sufficient amount of data a detailed measurement of the gain of each bar can be performed, details of this will be the subject of an upcoming paper by the collaboration. 

\begin{figure}[h!]
  \centering
    \includegraphics[width=9 cm]{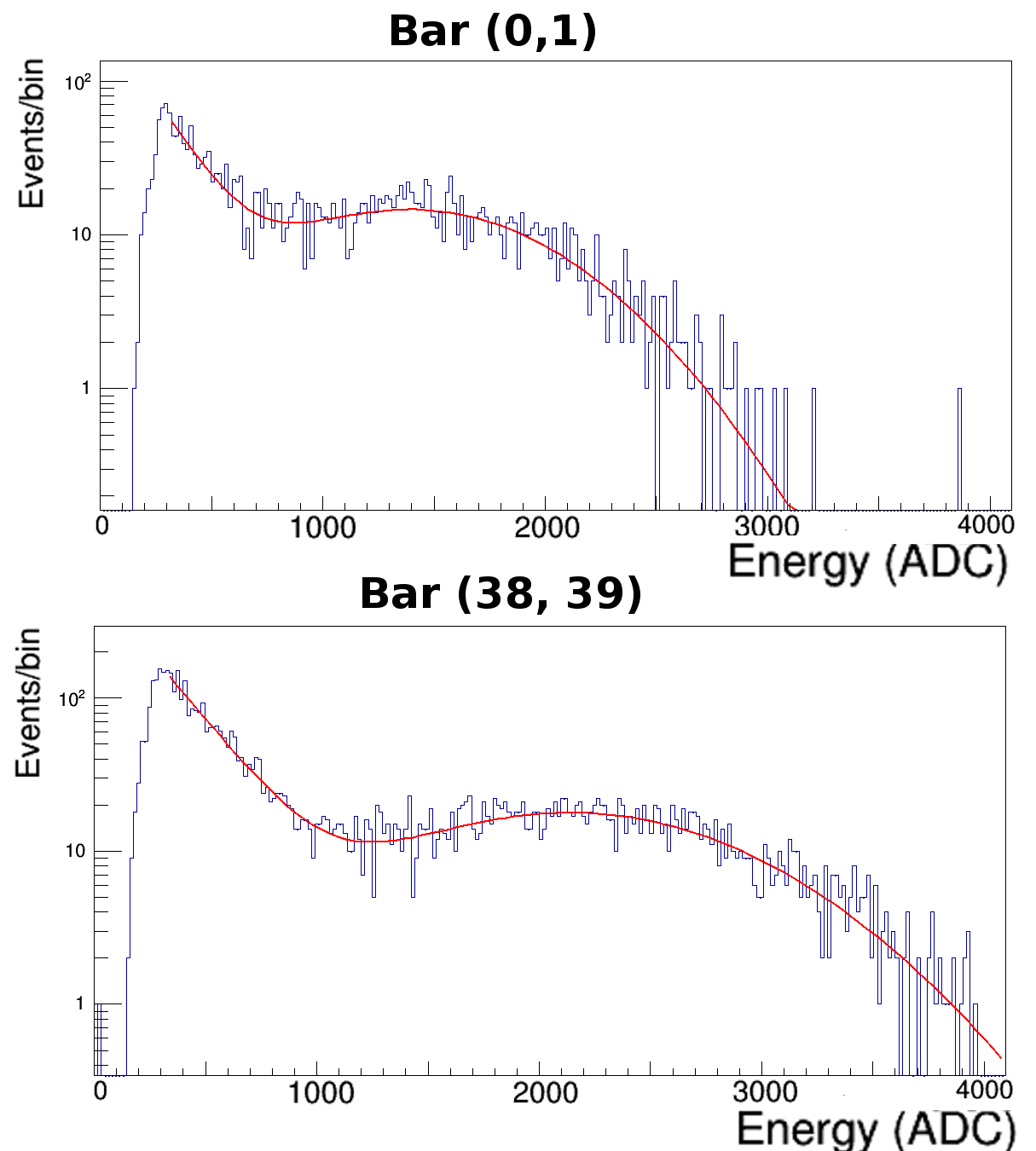}
  \caption{Spectra of two bars on different sides of the instrument, obtained from an 80 keV irradiation, fitted by an exponential plus a Gaussian. The mean of the Gaussian is used as an indicator for the gain of the bar. The figure indicates the relatively high level of non-uniformity present in the instrument during the beam test.}
\label{fig:energy}
\end{figure}

During the beam test such sources were not present but rather the beam itself was used to calibrate each bar. Using the recorded positions of the x and y translators time periods were selected for each bar during which the beam was aimed directly at it. The energy spectrum recorded during this time interval was then used to measure the gain of each bar. For this purpose the on-axis data from the 80 keV irradiation test was used. For each bar the spectrum was fitted with an exponential function plus a Gaussian which was found to represent the spectra well. The mean of the Gaussian was taken as an indicator for the gain of the bar. It should be clarified here that although the Gaussian measured while the beam illuminates the bar roughly corresponds to the photopeak, they are not identical. The Gaussian is contaminated by events induced by photons which scatter first, either in dead material, or which scatter several times in different bars before being absorbed in the illuminated bar. The measured peak position is furthermore affected by Birks' effect \cite{Birks} and is therefore not equal to the beam energy. These effects are however equal for all bars and the peak position, measured in visible keV, can therefore be used to calibrate the different bars. Examples of several fitted spectra are shown in figure \ref{fig:energy}. Bars are numbered based their number along the the x and y axis, for example, bar (1,0) is the second bar along the x-axis and the first along the y-axis. It should be noted here that prior to the beam test the high voltage settings of the different DMs were not fully optimized for a high uniformity within the full polarimeter. This was due to technical problems prior to the beam test. As a result the difference in gain between different DMs is relatively high compared to in-orbit data for which the high voltage settings were optimized. Furthermore for several DMs the HV was set slightly too low to perform an accurate gain measurement. As a result the uncertainty for the energy calibration of some bars is relatively large compared to what can be expected from flight data. This uncertainty in the energy calibration results in a systematic error on the polarization measurements as will be discussed later in this paper.

\subsection{Analysis Chain}
\label{sec:chain}

\begin{figure}[h!]
  \centering
    \includegraphics[width=15 cm]{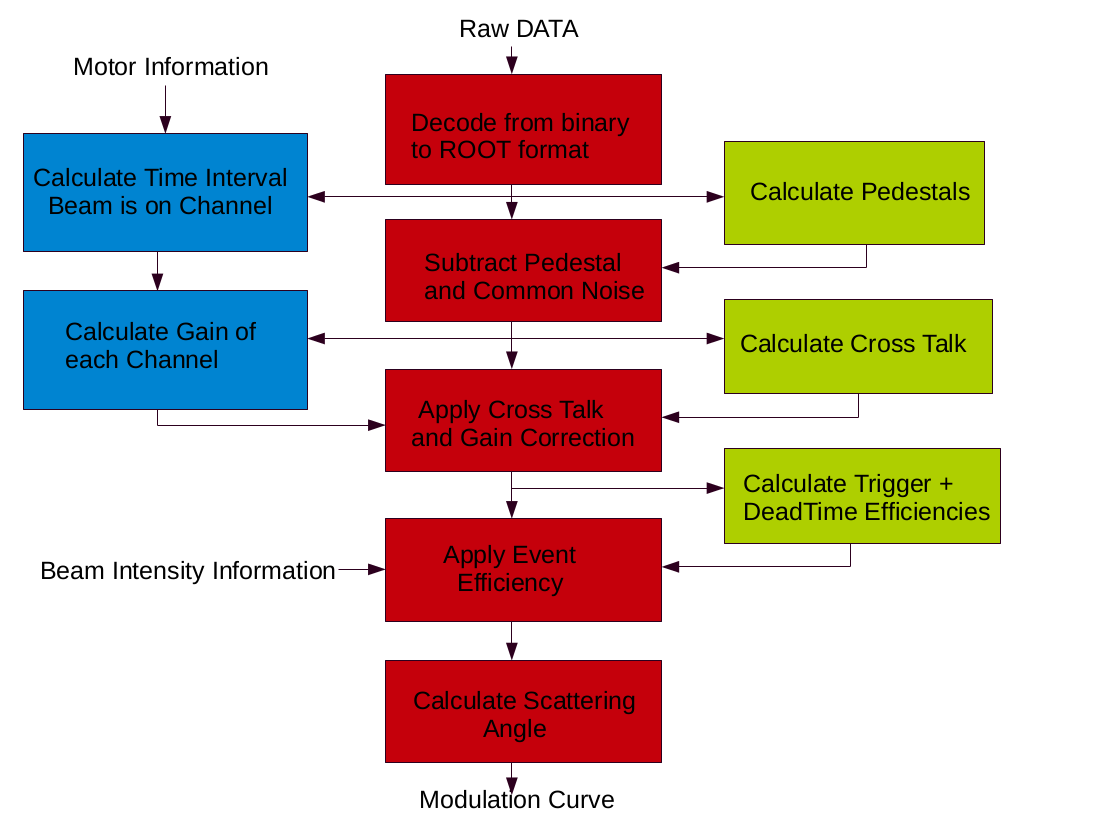}
  \caption{The flow chart of the data analysis chain as applied to ESRF data.}
\label{fig:flow}
\end{figure}

The full analysis chain performed on the beam test data is shown in the flow chart presented in figure \ref{fig:flow}. The process starts by decoding the data from binary format to ROOT format \cite{ROOT}. For each scan a time period covering the period from the beam entering the first bar and leaving the last bar is then selected. Subsequently the electronic pedestal levels as well as the noise of each bar are calculated. In POLAR each bar has a pedestal level which forms the lower limit of the analogue energy value, the upper limit is 4095 ADC. In order to measure the pedestal level a 'pedestal event' is taken every second. In such an event all bars in the instrument are triggered and read-out. The mean of the ADC level measured for each bar is the pedestal value while the spread in the pedestal level gives the noise. The noise in POLAR consists of two components, the common-noise, which is shared between all bars inside a DM and the intrinsic noise which is individual for each bar inside the DM. 

\begin{figure}[h!]
  \centering
    \includegraphics[width=15 cm]{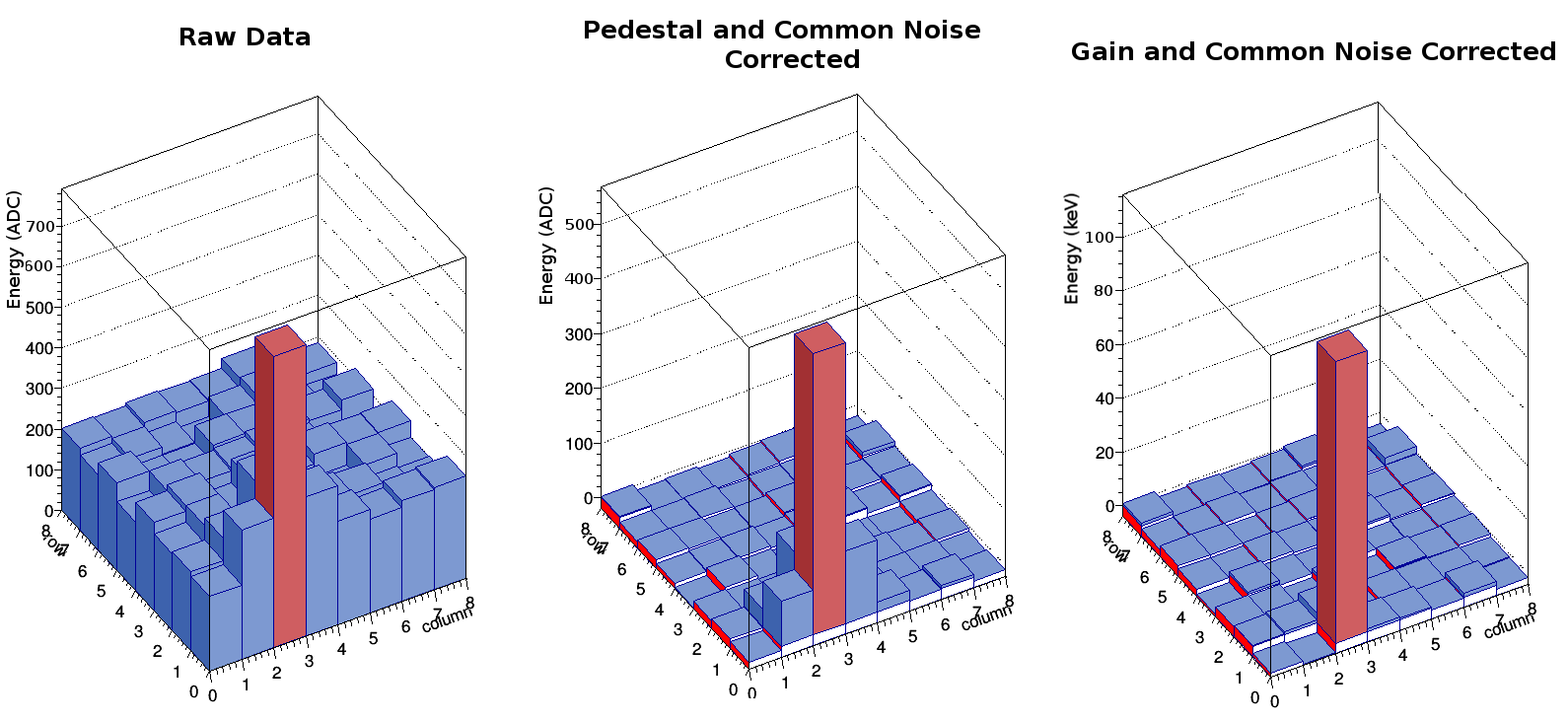}
  \caption{Example of an event inside a DM in the three first stages of the analysis. The raw data is shown on the left, in the middle the same event is shown after pedestal and common noise correction. Finally on the right the event is shown after crosstalk and gain correction.}
\label{fig:Event_disp}
\end{figure}

After calculating the pedestal levels the mean values are subtracted from each event together with the mean of the common noise, called the common shift level. The common shift level of each event is calculated by taking the mean deviation from the pedestal level of all the non-triggering bars. Triggering bars are excluded from the calculation as these likely have an energy deposition in them. The common shift level of each event is subtracted from all the bars in the DM. The remaining irreducible noise on each bar consists of the intrinsic noise together with the uncertainty on the common shift which is roughly $1/\sqrt{64}$ of this level. Typical levels for the intrinsic and common shift level are $10\,\mathrm{ADC}$ and $40\,\mathrm{ADC}$ respectively. A typical event inside a DM before and after pedestal and common noise correction is shown in the left and middle panel of figure \ref{fig:Event_disp}.

\begin{figure}[h!]
  \centering
    \includegraphics[width=9 cm]{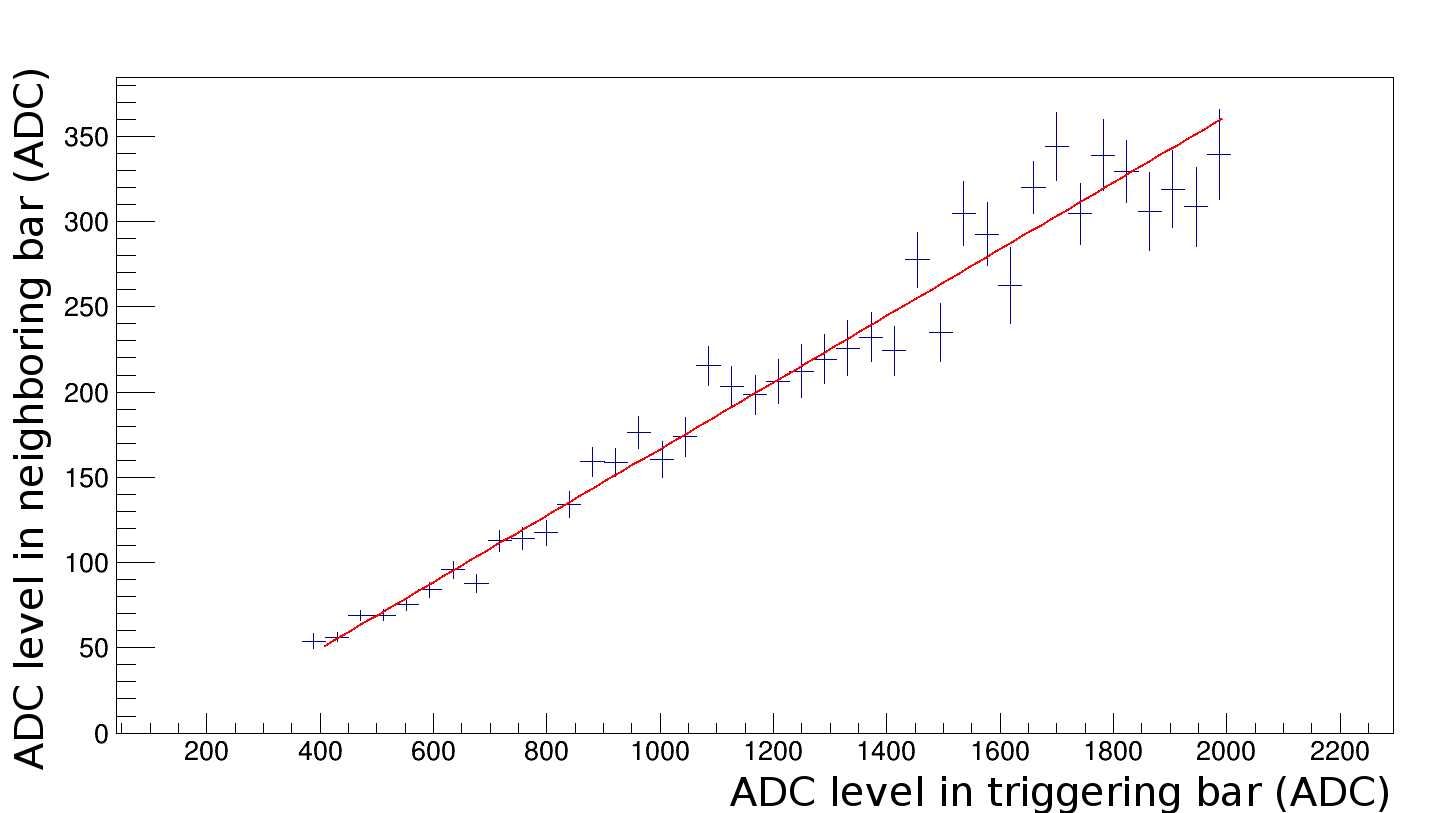}
  \caption{Example of the ADC energy deposition of a triggering bar on the x-axis versus the ADC level measured in a neighboring bar on the y-axis. The measured, non-gain corrected, crosstalk value for these two bars is taken as the slope of the fitted line which here is $19\%$.}
\label{Xtalk}
\end{figure}

Subsequent to pedestal and common noise correction the energy depositions as measured in ADC are converted both for gain, measured using the procedure described in the previous subsection, and for the crosstalk between bars. The crosstalk in POLAR is mostly a result of the use of the MAPMTs in combination with an optical coupler placed between the scintillator bars and the MAPMT. The MAPMT consists of an array of $8\times8$ anodes with a pitch of 6.08 mm and a single photo-cathode window which covers all the anodes. Optical photons exiting a scintillator bar non-parallel to the z-axis can produce photo-electrons which enter the anode of a neighboring bar on the MAPMT, thereby forming one crosstalk component. In POLAR this effect is enhanced by the requirement to place an optical coupler which both improves light collection and protects the MAPMT from vibrations. A second, less significant, component comes from electrical crosstalk within the MAPMT described in more detail for similar MAPMTs in \cite{Calvi} and \cite{Hualin}. A potential third, but significantly less important component of crosstalk which cannot be excluded stems from potential imperfections in the optical insulations of the scintillator bars. As most of the crosstalk occurs between the scintillator bars and the MAPMT and the DMs are optically isolated for each other, the crosstalk in POLAR is only present within a single DM. The presence of electrical crosstalk in the FEE was excluded by injecting known charges into each bar on the FEE and measuring the response in neighboring bars. The level of the crosstalk is measured by dividing the energy (measured in ADC) deposited in a triggering bar by that found in the neighboring bar. This is performed for all the bars within a DM for each event. The energy deposited in a neighboring bar can be plotted against that measured in a triggering bar resulting in a linear distribution, an example of which is shown in figure \ref{Xtalk}. The slope of the distribution shown in figure \ref{Xtalk} is taken as the crosstalk ratio.

As described in the previous section the gain of each bar is calculated by fitting the pedestal and common shift subtracted spectra for the time interval during which the beam was pointing directly at the bar. The time interval is selected using both the information on the motor positions, which gives 1 second precision, and by finding sudden changes in the rate in a bar which indicate the beam entering or leaving the bar giving a precision of about $0.05$ seconds. Using only data acquired during this time interval ensures that the photopeak is clearly visible in the spectrum and ensures that all bars are calibrated using similar data.

As the crosstalk is derived from energy depositions measured in ADC the measured crosstalk value needs to be corrected for the gain differences between the two bars. The gain and crosstalk correction are therefore performed at the same time using a procedure described in \cite{Hualin}. Application of these first three steps on a single event, induced by a single photon, in a DM in different stages of the analysis chain is shown in figure \ref{fig:Event_disp}. The final step shows the deposited energy in each bar in the DM corrected for pedestals, common noise and cross talk and converted to visible keV.

After the crosstalk and gain correction three different weights are applied to each event to remove systematic effects. The first is required to remove the variations in beam intensity during a run. For this purpose the recorded beam current is used to weight each event. Secondly the dead time measured during a scan will vary. The dead time is measured and recorded and is used as a second weight. Lastly each bar in POLAR has a different threshold position. The threshold position of each bar is set using two variables. The first is a threshold value set to each DM, the second is an offset value which can be applied to each bar within a DM. By optimizing both these values a relatively high level of uniformity in the threshold positions can be obtained. For the beam test at ESRF the optimization of the threshold position was not fully finished and as a result the spread in threshold positions during the beam test is significantly higher than that measured during in-orbit data taking. The ESRF beam test therefore does not serve as a best case scenario.

\begin{figure}[h!]
  \centering
    \includegraphics[width=9 cm]{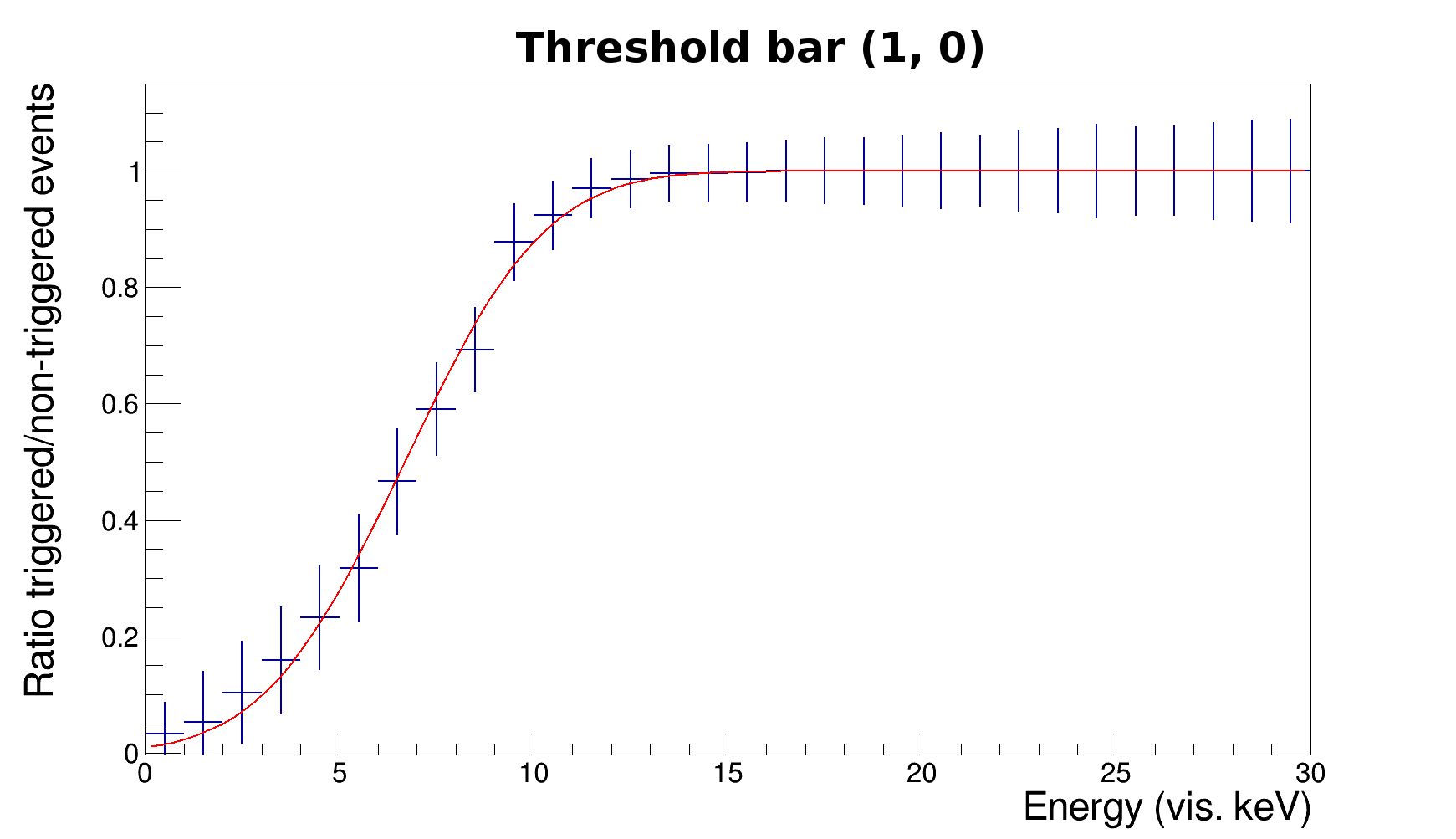}
  \caption{Example of a threshold measurement for a bar in POLAR. The x-axis shows the energy after crosstalk and gain correction while the y-axis shows the ratio of the number of triggered events over the number of all events.}
\label{fig:threhsold}
\end{figure}

The difference in the threshold position between bars results in a significant difference in detection efficiency between bars. In order to correct for this effect a third weight is assigned to each event based on the threshold positions. For this the threshold position of each bar in POLAR is first measured in visible keV. The threshold position is measured by dividing a spectrum, as measured in visible keV, only consisting of triggered events by the raw spectrum containing also non-triggering energy depositions, again measured in visible keV. The result, an example of which is shown in figure \ref{fig:threhsold}, can be fitted with an error function the mean of which is taken as the threshold position of the bar and the width of the error function as the threshold width. This method is applied on each bar and the median of all the  threshold positions is taken as a cut value. For the data taken at ESRF the median was found to be $10.9\,\mathrm{keV}$ which corresponds to approximately 6 photo-electrons.

For each event the two bars with the highest energy depositions are assumed to be the locations of the two interactions of the photon. The weight of the event is based on the threshold positions of these two bars. If the energy deposition in one of the two bars is below the median of all the thresholds the event is given a weight of 0, thereby removing the event from the analysis. For all other events a weight is calculated using:

\[\mathrm{weight} = \left[1 + \erf \left( \frac{\mathrm{E}-\mathrm{mean}}{\mathrm{width}}\right) \right]/2\]

\begin{figure}[h!]
  \centering
    \includegraphics[width=10 cm]{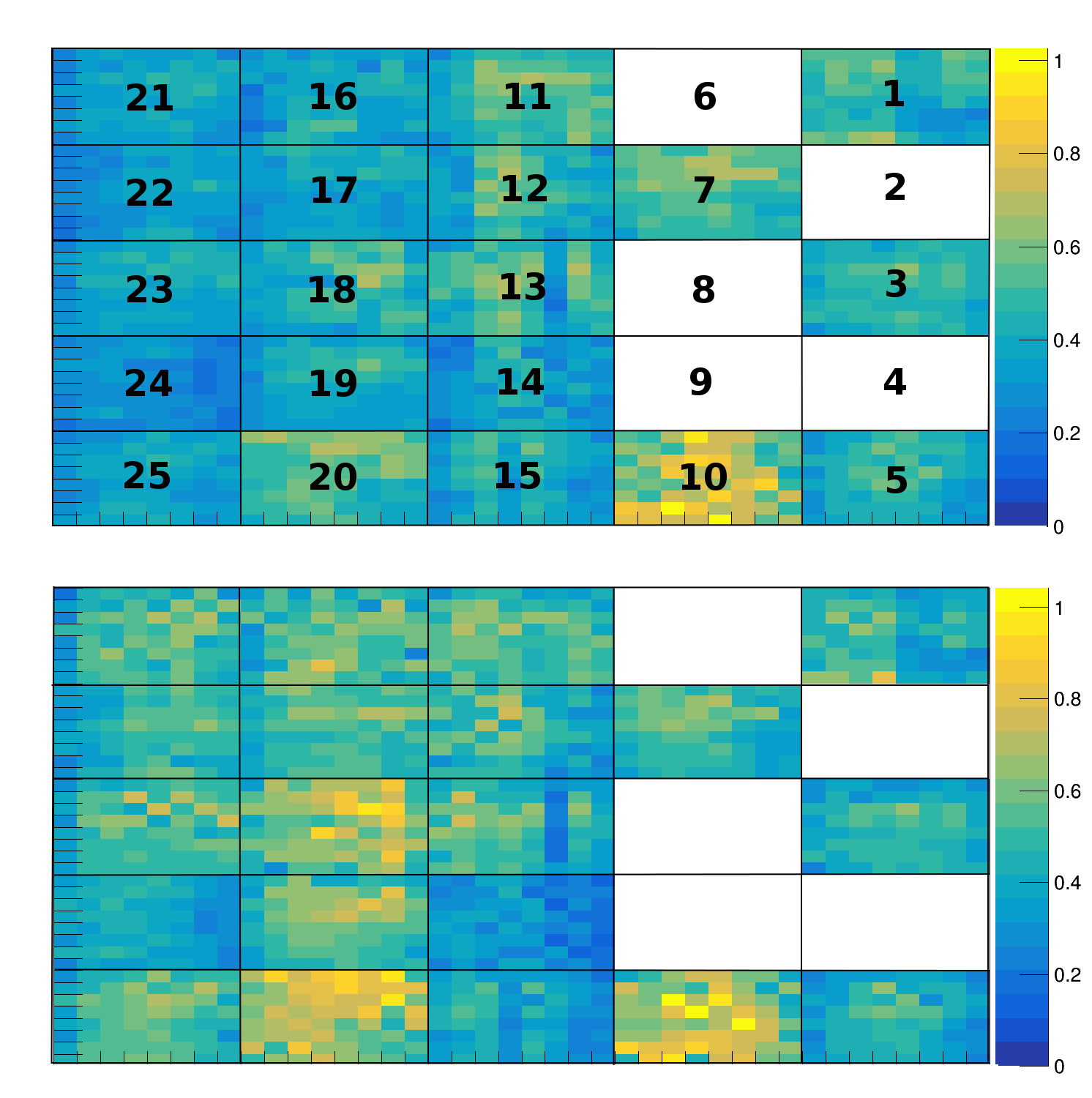}
  \caption{Maps of the measured trigger rates, normalized by the highest rate before (top) and after (bottom) weight applications. The numbers in the top figure indicate the DM numbering used in the analysis while the white spaces indicate DMs which are excluded from the analysis for this particular data set.}
\label{weights}
\end{figure}

Here E is the measured energy deposition, $\mathrm{mean}$ is the mean of the threshold position of the triggering bar and $\mathrm{width}$ is the threshold width of this bar. The weights of the two bars are then multiplied to give the final threshold based weight factor. The weighted event rate from bars with a low threshold will thereby decrease while that from bars with a higher threshold will increase. This procedure can therefore be seen as applying an offline smoothing of the efficiency of all the bars in order to produce a higher level of uniformity. The final total weight of an event consists of the multiplication of the dead time based, current based and threshold based weights. An example of the trigger rates, normalized by value in the bar with the highest trigger rate before and after applying the weights can be seen in figure \ref{weights}. It can be observed that the uniformity has been improved and a gradient (from left to right), induced by a difference in beam intensity, has disappeared. However, it can also be observed that the lack of high voltage calibration results in a clear difference between different DMs while differences between bars within DMs are retained, also after weight application, resulting from a lack of fine-tuning of the threshold positions. Lastly it should be noted that several DMs do not show data in this scan, which indicates the removal of several DMs in the analysis due to either FW or FEE issues.

After applying a weight to each event a modulation curve can be produced for each full scan of the instrument. The event selection criteria are based on that used in previous POLAR calibration campaigns, see \cite{Silvio} for details. First at least two bars are required to have a trigger in the instrument. The two bars with the highest energy deposition are selected. The position resolution is limited by the size of the scintillators. Again, as in the analysis presented in \cite{Silvio}, to smooth the final modulation curves a random position in the x-y plane is selected within the two participating bars to calculate the scattering angle which is accumulated into the modulation curve.

For the analysis presented here a second requirement is placed that the two bars participating to the event are not adjacent. If the two highest energy depositions within an event are adjacent the second highest energy deposition is ignored and the highest and third highest energy deposition are selected etc. It should be noted here that the event selection used here is largely based on that presented in \cite{Silvio} and can be further optimized. Detailed studies on the event selection, cut values and analysis methods for in-orbit data will be the subject of a future publication.

\section{Monte Carlo Simulations}

\subsection{Geant4 Simulations}

\begin{figure}[h!]
  \centering
    \includegraphics[width=10 cm]{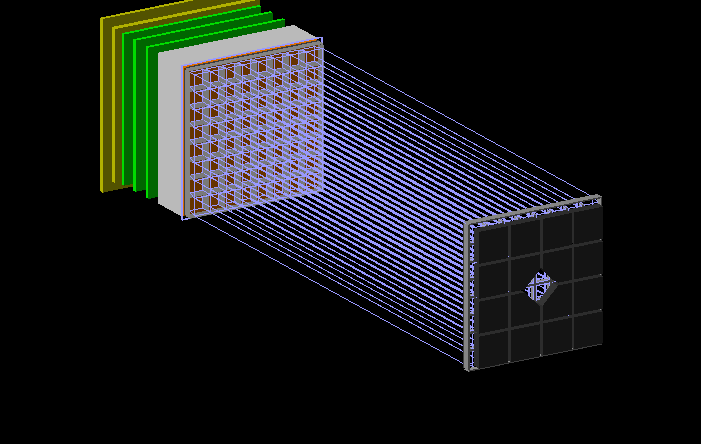}
  \caption{A DM as used in the simulations, the shielding was removed to make the different parts visible. The DM consists of the 64 scintillators (transparent blue), covered by rubber dampers (black). At the bottom the PMT (gray), the optical coupling (orange), the electronics (green) and the mechanical coupling (yellow) can be seen in the figure.}
\label{module_sim}
\end{figure}

The Monte Carlo package used for simulating the POLAR detector makes use of Geant4.10.2 \cite{G4}. The package, which was developed as a dedicated continuation of simulation packages previously used by the collaboration \cite{Shaolin}\cite{Estela}, contains a detailed detector model. An example of a simulated DM can be seen in figure \ref{module_sim}. The package uses the G4LivermorePolarizedComptonModel which includes an accurate modeling of polarized Compton scattering. Simulations of the beam test were performed by simulating the irradiation of the instrument using 40 rectangular bars of $0.6\times300\,\mathrm{mm}$ each of which is centered on one row of bars. The simulations therefore simplify the scanning of the instrument by instead having an instantaneous irradiation of all bars, however, the irradiated area is equal to that during the measurement. As a result the change in the beam current is not taken into account in the simulations but is rather only corrected for in the measurement data. Apart from the lack of application of the current based weighting factor as well as that of the dead time based weighting factor the output of the simulations proceeds equal to the analysis of the measurement data. 

The Geant4 package of POLAR only handles the physical interactions inside the instrument. This includes a correction for the non-linear light yield due to the Birks' effect \cite{Birks}. The Birks' constant used in the simulations is taken from \cite{Radek} in which this value is measured for the scintillation material used in POLAR. The package outputs a ROOT file containing the energy, corrected for the Birks' effect, and the position of each interaction in the scintillator volume. This output is then handled by a second ROOT based program which handles the digitization part, this is described in the following section. The output of the digitization is a ROOT file with the same structure as the data produced by the instrument and can therefore be handled by the same analysis software.

\subsection{Digitization}

Below is a description of the different steps of the event digitization. An event is defined here as all interactions within a coincidence window, for simulations this implies all the interactions induced by one primary particle. Both normal and pedestal events, events where all bars are artificially triggered and read out, are created in the digitization part. The pedestal events finish already after step 1, while normal events follow all the below mentioned steps.

1) Every event starts with the addition of the pedestal level to each bar. This is followed by adding both the intrinsic noise and the common noise for each bar. The values of the intrinsic noise, common noise and the pedestal are taken from measurement data. The pedestal is added as a constant, the intrinsic noise and common noise are taken from a Gaussian distribution centered around 0 with the width taken from data. For the common noise it should furthermore be noted that the level of this noise has a strong temperature dependence, for this purpose the temperature can be given as an input to the program and the measured relation between common noise and temperature is used to provide the common noise width. Pedestal events are also created in the simulations. Each data run starts by default with 1000 pedestal events, similar to the data. This is followed by 1 pedestal event being taken every second.

2) Every event handled by Geant4 results in a list consisting of all the interactions in the scintillator volumes. For each interaction the x,y and z positions of the interaction are given together with the visible part of the deposited energy, meaning the deposited energy corrected for the Birks' effect.

3)For each interaction within an event the x, y and z positions and energy deposition are read in. Based on the x and y positions the scintillator bar of the interaction is determined. The z-position is required as the number of optical photons reaching the MAPMT depends on the interaction position within the scintillator, this effect therefore needs to be handled in the simulation as well. It was decided to do this in the digitization part of the simulations. In order to study the dependency of the optical yield on the z-position optical simulations were performed with Geant4. Through these simulations it was found that, with exception of the top and bottom $4\,\mathrm{mm}$ of the scintillator, the optical yield is independent on the position and was found to be $(32\pm2)\%$. Meaning $32\%$ of the optical photons produced reaches the MAPMT while the other $68\%$ is absorbed either in the scintillator, the wrapping material or the medium between the two. In the top and bottom $4\,\mathrm{mm}$ of the scintillator, where the scintillator is conically shaped the optical yield is significantly higher and increases linearly to $(60\pm3)\%$ at the top and bottom edge of the scintillator.

\begin{figure}[h!]
  \centering
    \includegraphics[width=11 cm]{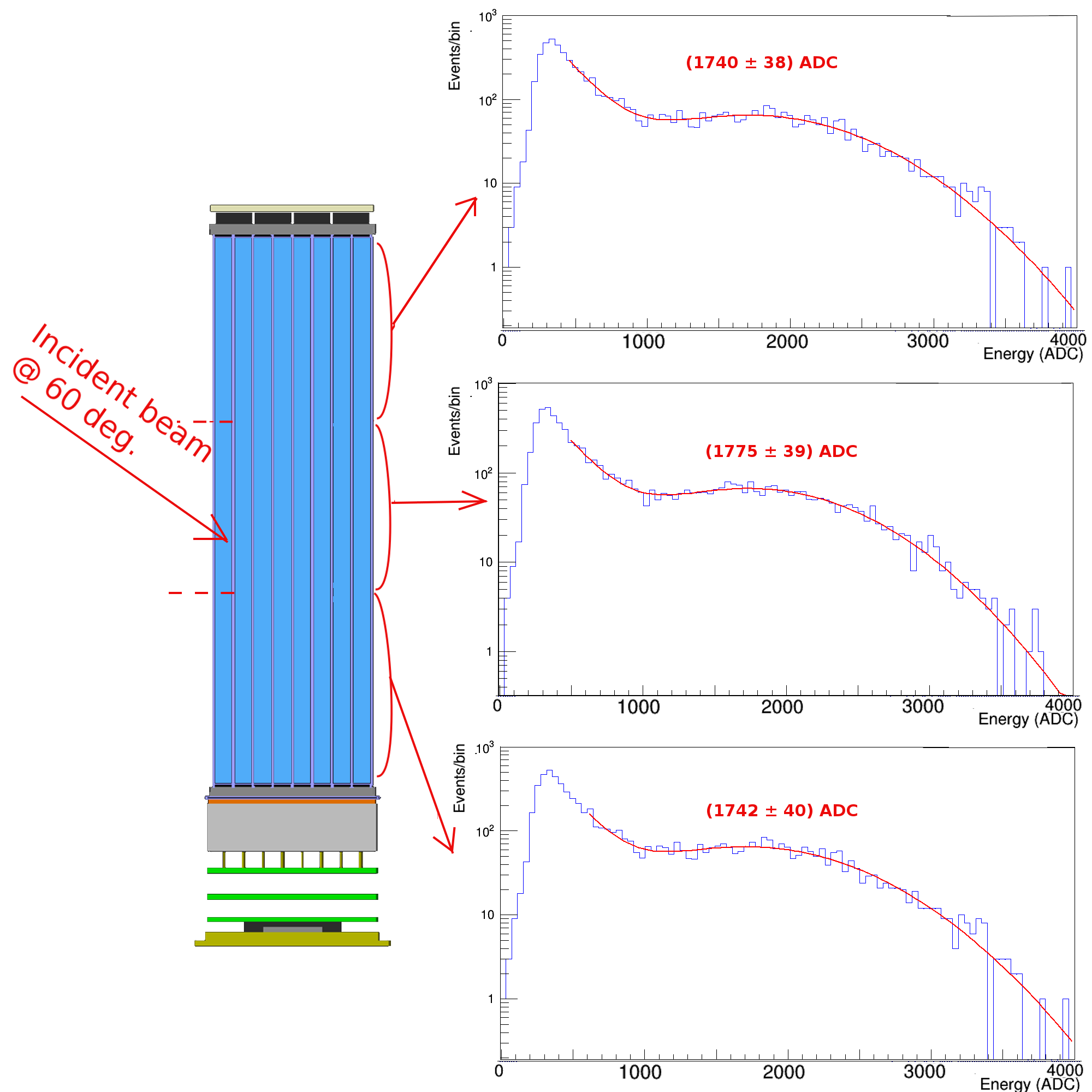}
  \caption{The spectra as measured by the bar on the left in the schematic drawing for different periods during an off-axis scan with an 80 keV beam. The top figure shows the spectrum as measured while the top third of the bar is illuminated, the middle shows the spectrum while the middle third is illuminated and the bottom when the bottom is illuminated.}
\label{pos_dep}
\end{figure}

The lack of dependency of the light yield in the majority of the scintillator was measured as well using data from the off-axis irradiation measurements at ESRF. Data taken from different periods while the beam was directly illuminating a scintillator bar positioned on the outside of the instrument, which was at an angle of $60^\circ$ with respect to the beam, was taken to produce three different energy spectra. The three different spectra, taken for the periods where the beam was scanning the top, middle and bottom third of the bar are shown in figure \ref{pos_dep}, the mean position of the photo-absorption peak in the three plots are $1740\pm38$ ADC, $1775\pm39$ ADC and $1742\pm40$ ADC respectively. From this measurement, as well as others performed on other bars, we can conclude that the light yield dependence on the interaction position within a scintillator bar is negligible in the majority of the bar. For the simulations the constant value of $32\%$ is therefore used for the bulk of the scintillator while in the conical ends of the scintillator the light yield linearly increases to a maximum of $60\%$ at the edge, based on the performed optical simulations.

4) The energy deposition is converted into a number of optical photons. For the the light yield a value of 9.2 optical photons per keV is taken from the EJ-248M data sheets \cite{EJ248M}. The final number of optical photons is then picked from a Gaussian distribution.

5) The optical cross talk of each bar to all other bars within the DM is taken from the measurement data and used as input to the simulations. Based on this data a percentage of the optical photons, which differs for each bar, is moved from the bar to the neighboring bars. The number of photons escaping to the neighbors follows a Poisson distribution. For each bar with an energy deposition a random number is picked from the Poisson distribution to calculate the number of optical photons which are moved to each neighboring bar in the DM. For each neighboring bar the Poisson distribution is based on the optical cross talk value from the data. The total number of photons escaping to the neighbors is added up and subtracted from the bar with the energy deposition.

6) For each bar with an energy deposition the number of photons is now converted to photo-electrons using the quantum efficiency of $20\%$ taken from the Hamamatsu data sheet \cite{Ham}. The number of photo-electrons is taken from a Poisson distribution.

7) The number of photo-electrons is converted to an electrical signal using a Gaussian distribution centered around the number of photo-electrons with a relative width of 0.6. This value is taken from the measurements of the single photo-electron peak.

8) The electrical signal is converted to ADC using the gain values from measured data. The gain values have been measured for each bar, the gain of each bar in the simulations therefore matches that of the real data. These gain values have to be corrected for crosstalk through a process described in appendix A. Furthermore the FEE gain shows small deviations from a perfectly linear gain at ADC values below 1000 ADC. These deviations can be measured using an internal calibration mode of the FEE which makes use of a known charge injected into the FEE using a capacitor. The measured deviations are taken into account in the simulations.

9) All the energy depositions in one event are added up together. Energy depositions in POLAR are stored as 12 bits, and therefore overflow appears above 4095 ADC. Depositions exceeding this value are therefore set to 4095.

10) The trigger threshold is applied to every event. For each bar the mean value of the threshold position and the width of the threshold are read-in from a Gaussian deduced from measurement data. The threshold position and width of each bar therefore matches that in the real data. A random value is picked from the Gaussian and compared to the energy deposition in the bar to determine if the bar triggers or not.

11) For each bar a check is performed whether it has triggered. Triggered bars are flagged as such and the total number of triggers within each DM is counted. DMs without a trigger are not read-out. A counter is used to store events with one trigger which are read-out based on the prescaling value described in section \ref{sec:1}. The same holds for events with too many triggers. During the beam test the maximum multiplicity was set rather low at its default value corresponding to approximately 4 triggered bars. 

12) Events are finally written to a ROOT file with the same structure as that used in real data.

\begin{figure}[h!]
  \centering
    \includegraphics[width=10 cm]{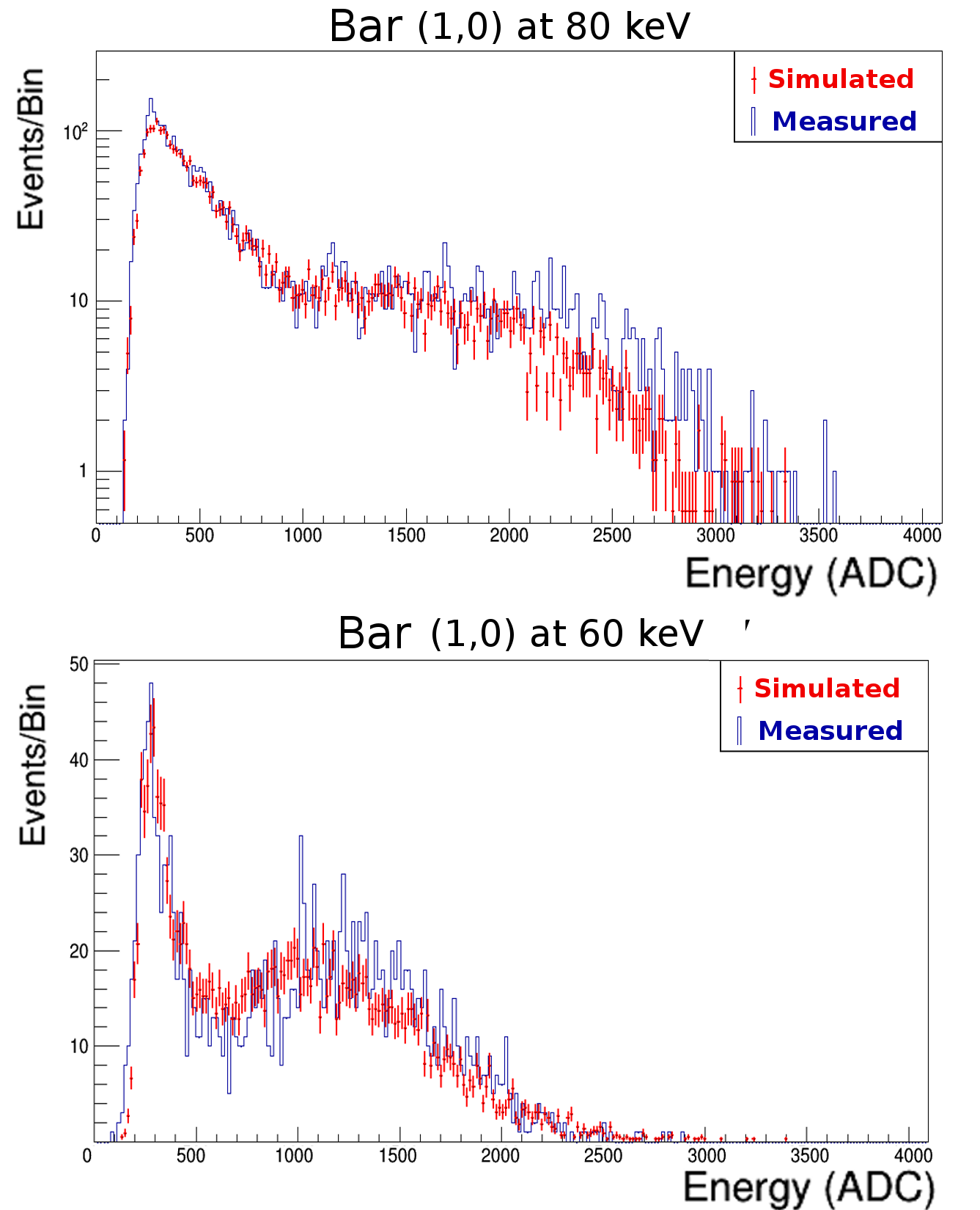}
  \caption{Comparison between spectra from measurements (histogram) and simulations (crosses) produced in a single bar in POLAR for two different energies for bar (1,0), 80 keV (up) and 60 keV (down).}
\label{fig:sim_meas}
\end{figure}

\begin{figure}[h!]
  \centering
    \includegraphics[width=10 cm]{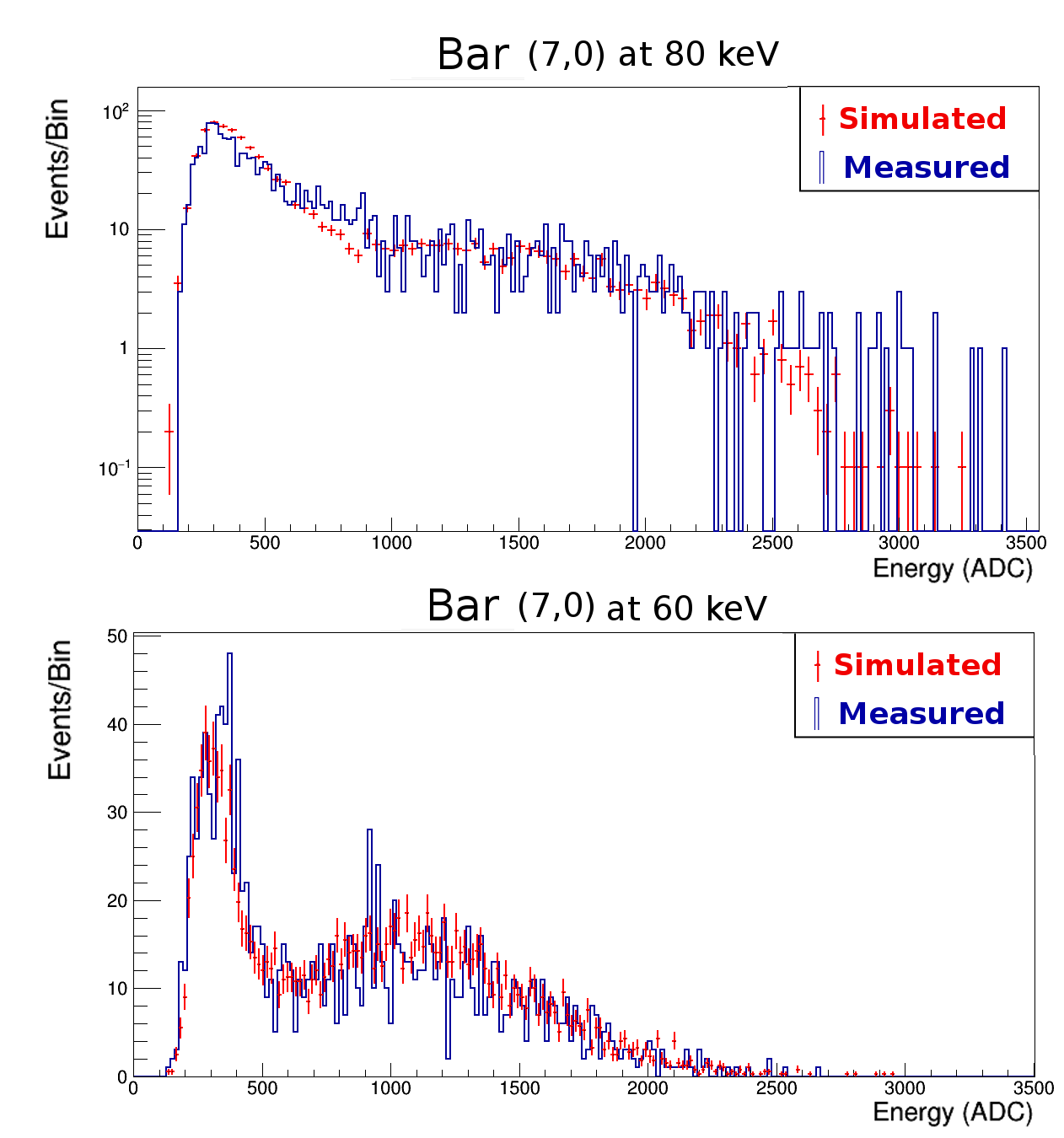}
  \caption{Comparison between spectra from measurements (histogram) and simulations (crosses) produced in a single bar in POLAR for two different energies for bar (7,0), 80 keV (up) and 60 keV (down).}
\label{fig:sim_meas2}
\end{figure}

In order to perform a first verification of the simulation procedure the spectra produced can be compared with the measured data. For this purpose simulations were performed where only one bar was irradiated by a rectangle of $0.6\times6\,\mathrm{mm^2}$ equal to an irradiation of a bar when the beam is moving over it. These spectra can then be directly compared with the measured data for events selected with the beam directly irradiating this bar. An example of two comparisons is shown in figures \ref{fig:sim_meas} and \ref{fig:sim_meas2}. The figures, in which the simulated data is scaled to match the number of events in the measured spectrum can be seen to match the spectral shape over the full energy range quite well for both bars. It is however visible in figure \ref{fig:sim_meas} that the simulated gain is not as high as the measured gain while in figure \ref{fig:sim_meas2} the simulated and measured data match each other. As the simulated gain for each bar is dependent on both the crosstalk and the gain calibration of the neighboring bars a poor gain measurement in one bar can affect the simulated gain of the neighboring bar. The non-perfect simulated gain factors such as those in figure \ref{fig:sim_meas} are therefore a result of errors in the gain and crosstalk calibration of POLAR which, as mentioned earlier, was not optimum for the data used. 

\section{Response to Polarized Photons}

The measurement data for all irradiations for 60,80 and 110 keV beam was analyzed and has been compared to simulation data. As described in the simulation section the analysis used for both data samples is equal with the exception that dead time correction and beam current correction through a weighting method is not applied in the simulation data. The threshold level used for the analysis is $10.9\,\mathrm{keV}$. In this section the different error sources in the analysis and in the simulation are discussed first. After this the simulation and measurement results for the 110 keV beam data are compared in detail and the method for removing geometrical and instrumental effects from the modulation curves is tested. This is followed by an overview of the polarization results of the 60 and 80 keV irradiations.

\subsection{Systemic Errors}

The systematic error was determined both for measurement data and for simulation data. It was found that the systematic error on results from simulation data is dominant in this comparison. For the measurement data the main source of error stems from an error on the measurements of the gain values which are relatively large for the analyzed data. Although for most modules the measurement of the gain was possible with an error of the order of a few percent, for 3 of the modules taken into account in this analysis the error on the gain can be as large as $20\%$. This relatively large error is a result of an unclear photopeak, the mean of which can shift by up to $20\%$ depending on the fitting procedure. The error on the gain directly results as well in an additional error on the measured cross talk which, independent of the gain error, has an average measurement error of $10\%$. The error induced on the value of $\mu_{180}$, the amplitude of the 180 degree modulation, by an error of $20\%$ on the gain of all bars was studied by adding $20\%$ to the gain values used in the analysis chain, while also varying all cross talk values within a $10\%$ error margin,  and comparing the results with the normally processed data. It was found that for modulation curves the relative systematic error on $\mu_{180}$ induced by the error on the gain and cross talk for a single module is $0.1\%$. As in the data analyzed here 3 modules suffered from such bad energy calibration the relative systematic error resulting from the bad gain measurement in a polarized beam is $0.3\%$. 

Other possible factors such as errors on the threshold position, beam current and dead time were also studied. In the case of the beam current measurement the effect was studied by ignoring the weight from the beam current in the analysis, no significant differences in the measured $\mu_{180}$ values were found, showing that even when assuming a large error on the beam current measurement this will not induce any error on the measured value of $\mu_{180}$. Similarly the values of the thresholds were varied randomly for all bars within their relative measurement error, which is about $10\%$, and the resulting value of $\mu_{180}$ was again not found to change significantly in the results of the 10 times this analysis was repeated. The dead time correction follows a binomial distribution with a width of $2\%$. Applying a $2\%$ increase or decrease of the dead time in the analysis was again not found to affect the value of $\mu_{180}$ significantly. The error due to instrument alignment was studied by comparing the modulation resulting from simulations where the instrument was perfectly aligned and where beam was displaced by $0.5\,\mathrm{mm}$. Again no significant variation on $\mu_{180}$ were found. The typical relative systematic error on the measurement results of polarized data is therefore $0.3\%$ and results only from the error on the gain measurements of several of the modules. 

The systematic error on the simulated modulation resulting from the uncertainty of the gain values of several modules is significantly larger than that on the measured data. This is a result of the error now affecting the data during the analysis phase, similar to the effect on the measured data, as well as in the simulation phase. The relative systematic error on the $\mu_{180}$ value induced by a $20\%$ error on the gain of all bars and variations of the cross talk by $10\%$ in one module was found to be $1.1\%$. The total relative systematic error due to poor calibration of 3 modules was found to be $3.3\%$. All other possible factors which could contribute to the systematic error, such as uncertainties on the optical yield, the quantum efficiency, the resolution of the MAPMT and the threshold position were investigated by varying these values with reasonable uncertainty levels. It was found that the contribution of all such values was negligible. The systematic uncertainty in the simulations is therefore found to be mainly a result of the poor calibration of several modules. A similar approach was followed to get the systematic errors on the amplitude of the 360 degree modulation $\mu_{360}$, the amplitude of the 90 degree modulation $\mu_{90}$, and on the unpolarized samples, also there it was found that the systematic error is dominated by the poor calibration of several modules.

\subsection{110 keV}

\begin{figure}[h!]
  \centering
    \includegraphics[width=10 cm]{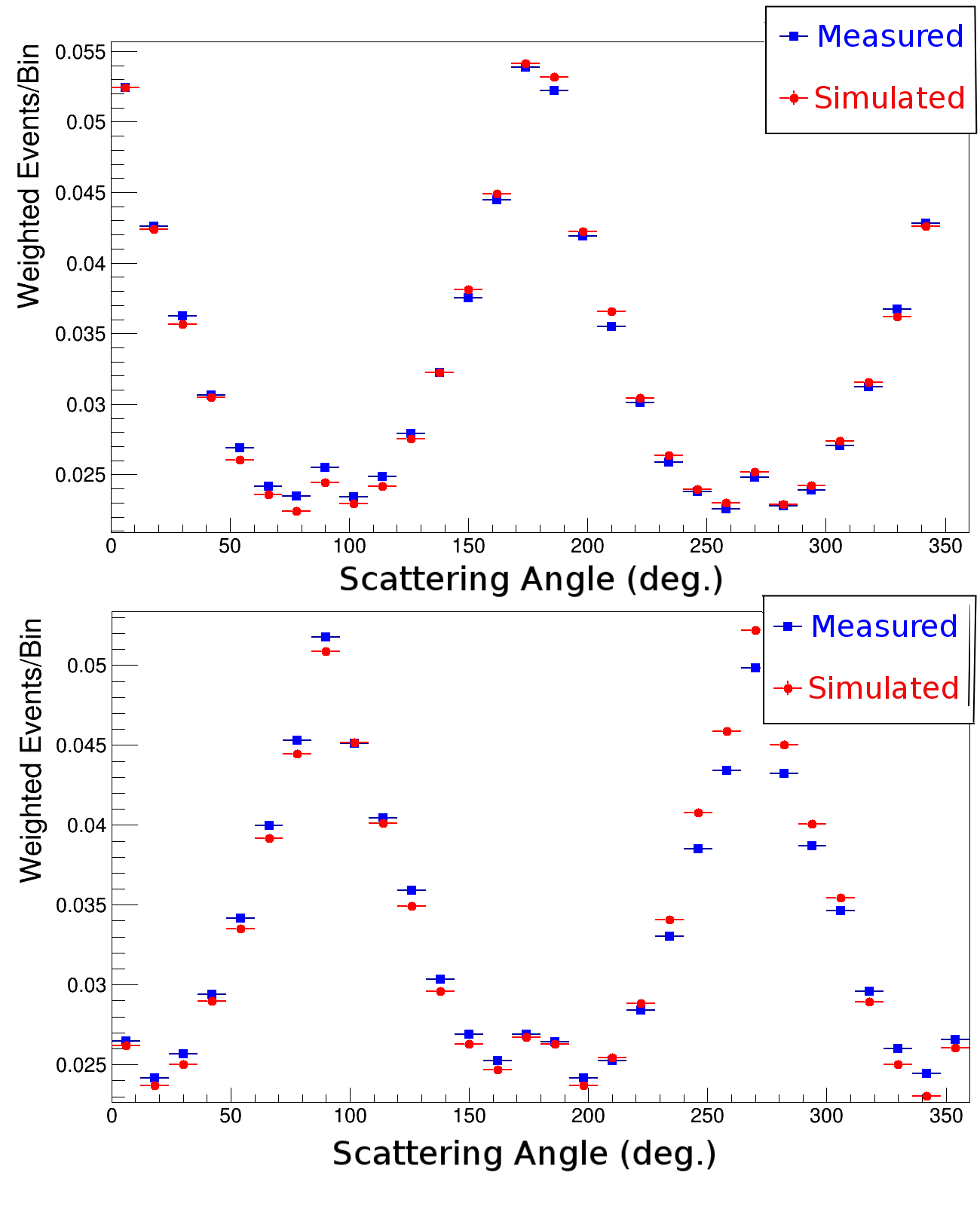}
  \caption{The measured polarized modulation curves (square) for two different polarization directions (top and bottom) using the full POLAR instrument, together with the simulated results (circles). The error bars only include the statistical error.}
\label{pol_110}
\end{figure}

Figure \ref{pol_110} shows the instruments response to $100\%$ polarized 110 keV photon beam for two different polarization directions both from simulated and from measured data. For the simulation data the same DMs were excluded in the analysis as the DMs which had to be excluded in the data due to either FEE or firmware issues. For the 110 keV data it concerns a total of 5 DMs, DMs 2,4,8,9 and 23 as indicated in figure \ref{weights}, which had to be excluded from analysis. For comparison purposes all 4 modulation curves were fitted with a function of the type:

\[A*(1+\mu_{360}*\cos(1*x+\phi_{360})+\mu_{180}*\cos(2*x+\phi_{180})+\mu_{90}*\cos(4*x+\phi_{90}))\]

In this function A gives the mean of the harmonic function, while $\mu_{360},\mu_{180}$ and $\mu_{90}$ give the amplitude of the $360^\circ$, $180^\circ$ and $90^\circ$ components of the modulation curves. The angles of the components are given by $\phi_{360},\,\phi_{180}$ and $\phi_{90}$ respectively. The amplitude values are summarized in tables \ref{table:110_curve1} and \ref{table:110_curve2} where both the fitting error and the systematic error are provided. The results show that the simulated and measured modulation curves are in good agreement. Although some differences are visible these are within the systematic errors dominated by the poor energy calibration of several DMs. It would be possible to improve the simulation results by tuning the gains of different bars for these to better fit the data. However, as such tuning is not possible for in-orbit data we chose to not do this here and use only methods we can apply also to space data. Furthermore as explained earlier the gain measurements in-orbit are expected to be significantly more precise due to higher statistics and more optimized settings of the instrument. 

\begin{figure}[h!]
  \centering
    \includegraphics[width=12 cm]{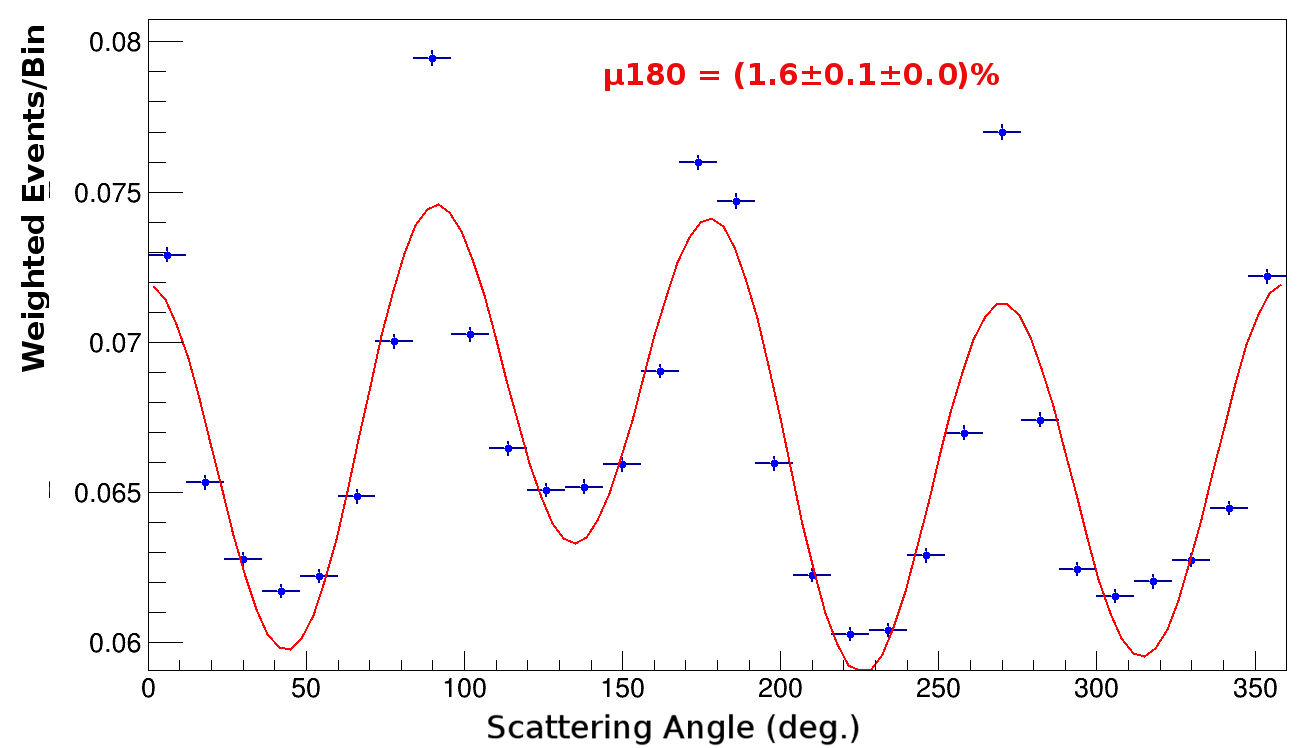}
  \caption{An example of an unpolarized modulation curve formed using the data from two measured different polarization angles. The fitted value of $\mu_{180}$ is provided along with the fitting and the systematic error, in that order. The data points only contain the statistical error.}
\label{unpol_meas}
\end{figure}

An unpolarized sample can be produced by adding the modulation curves from the two orthogonal polarization directions together. An example from the measured 110 keV on-axis beam data is shown in figure \ref{unpol_meas}. It can be seen that the remaining modulation for this sample is non-zero. This is partially a result of some DMs being excluded from the analysis and of the relatively poor high voltage calibration of the instrument which results in preferred scattering directions. The unpolarized sample can be used to remove the geometrical and instrumental effects from the polarized modulation curves. For this purpose the polarized modulation curves are divided by the unpolarized curve. This method was presented by the collaboration in \cite{Silvio}. 

\begin{table}
\begin{center}
\begin{tabular}{ |c|c|c| } 
 \hline
 Scan & Measurement Horizontal Scan ($\%$)  &  Simulated Horizontal Scan ($\%$) \\ 
 \hline
 $\mu_{360}$ & $1.5\pm0.1\pm0.1$ & $1.4\pm0.2\pm1.0$ \\ 
 $\mu_{180}$  & $39.4\pm0.1\pm0.1$ & $40.4\pm0.2\pm1.3$ \\ 
 $\mu_{90}$ & $12.2\pm0.1\pm0.1$ & $12.2\pm0.2\pm0.4$ \\ 
 \hline
\end{tabular}
\caption{Summary of the fit parameters found using a 110 keV on-axis beam while scanning the instrument which is in the horizontal direction, the fitting error and the systematic error are given in this order.}
\label{table:110_curve1}
\end{center}
\end{table}

\begin{table}
\begin{center}
\begin{tabular}{ |c|c|c| } 
 \hline
 Scan & Measurement Vertical Scan ($\%$) & Simulated Vertical Scan ($\%$) \\ 
 \hline
 $\mu_{360}$ & $2.4\pm0.1\pm0.1$ & $1.1\pm0.2\pm1.0$  \\ 
 $\mu_{180}$ & $33.0\pm0.1\pm0.1$ & $35.0\pm0.2\pm1.2$ \\ 
 $\mu_{90}$ & $10.3\pm0.1\pm0.1$ & $11.1\pm0.2\pm0.4$  \\ 
 \hline
\end{tabular}
\caption{Summary of the fit parameters found using a 110 keV on-axis beam while scanning the instrument which is in the vertical direction, the fitting error and the systematic error are given in this order.}
\label{table:110_curve2}
\end{center}
\end{table}

\begin{figure}[h!]
  \centering
    \includegraphics[width=10 cm]{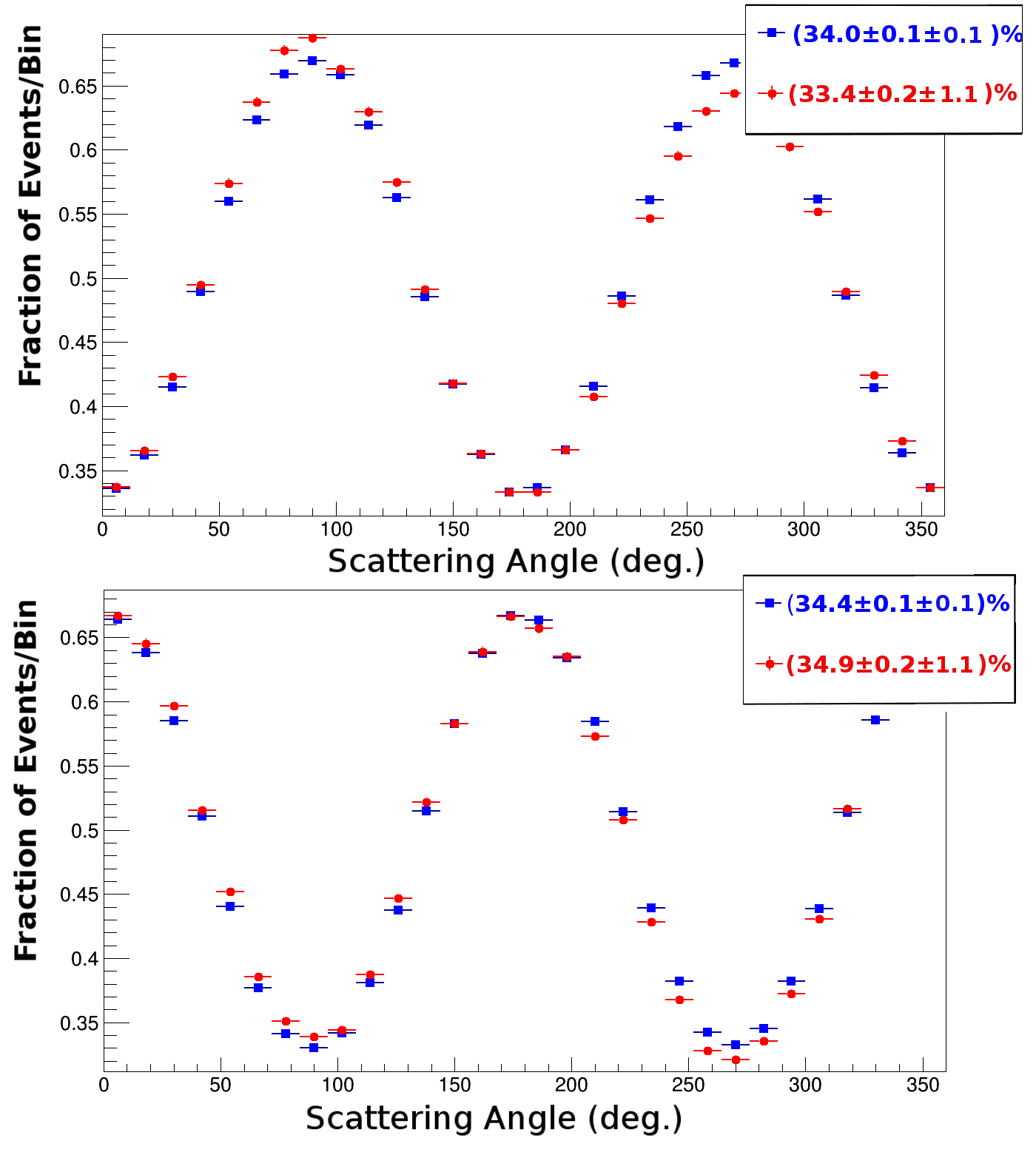}
  \caption{The measured polarized modulation curves for 110 keV divided by the measured unpolarized modulation curve (square) and the measured polarization curves divided by the simulated unpolarized sample (circle) for the full instrument for two different polarization directions (top and bottom). The data points only contain the statistical error. The fitted modulation factor is shown for each histogram.}
\label{110keV_corr}
\end{figure}

Two unpolarized samples were created, one from the measured data and one from the simulated data. When dividing the two measurement polarization curves by the unpolarized modulation curves the only significant component left in the modulation curve is the $180^\circ$ component. The corrected modulation curves, two corrected with measured and two with simulated data can be seen in figure \ref{110keV_corr}. The resulting $\mu_{100}$ value can now be acquired by fitting the data with the same harmonic function as used earlier on the non-corrected curves. From the pure data sample the results are $(34.0\pm0.1\pm0.1)\%$ and $(34.4\pm0.1\pm0.1)\%$ whereas from the data corrected using simulation the values are $(33.4\pm0.2\pm1.1)\%$ and $(34.9\pm0.2\pm1.1)\%$. The results acquired using both methods are therefore well compatible. Fitting these modulation curves with the same function but excluding the 360 and 90 degree component yields the same modulation factors.

%

%
%

\begin{figure}[h!]
  \centering
    \includegraphics[width=10 cm]{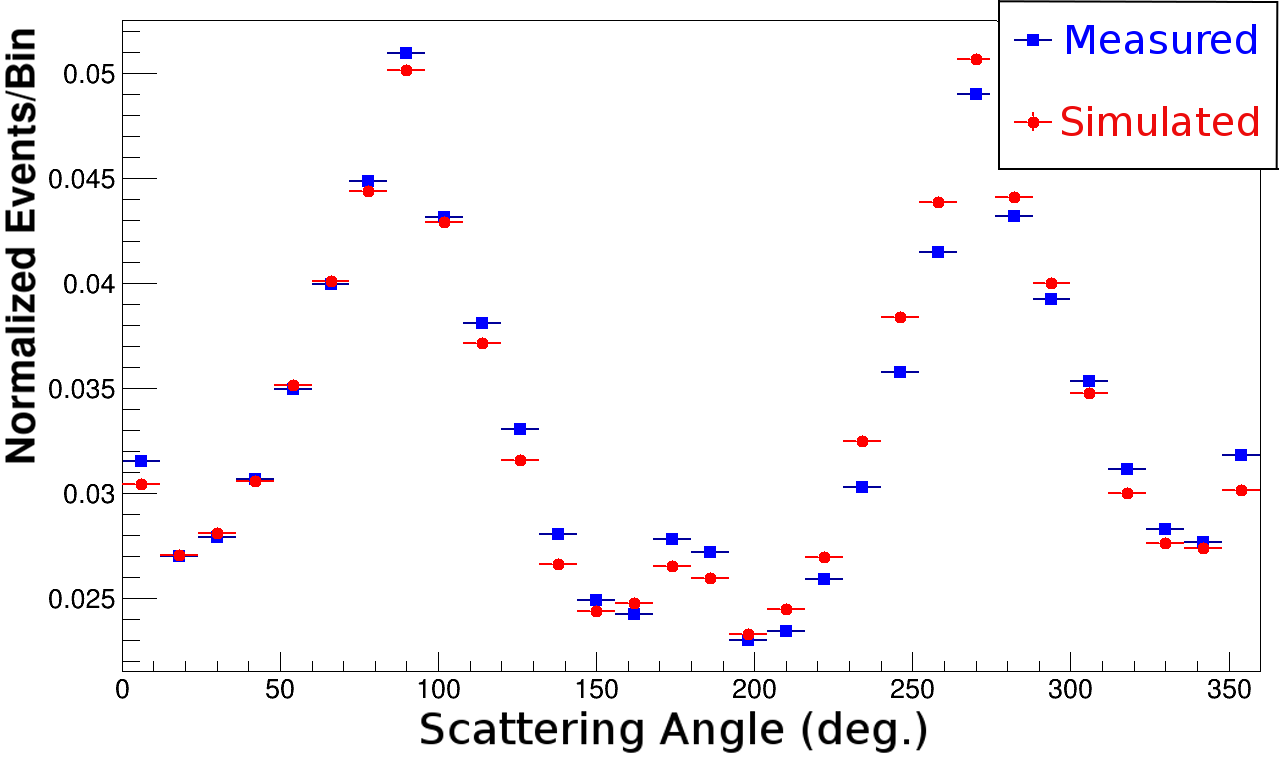}
  \caption{The measured polarized modulation curves for a 110 keV beam for a $30^\circ$ off-axis scan of the full detector from measurement (squares) and from simulation (circles).  The data points only contain the statistical error.}
\label{30}
\end{figure}

\begin{figure}[h!]
  \centering
    \includegraphics[width=10 cm]{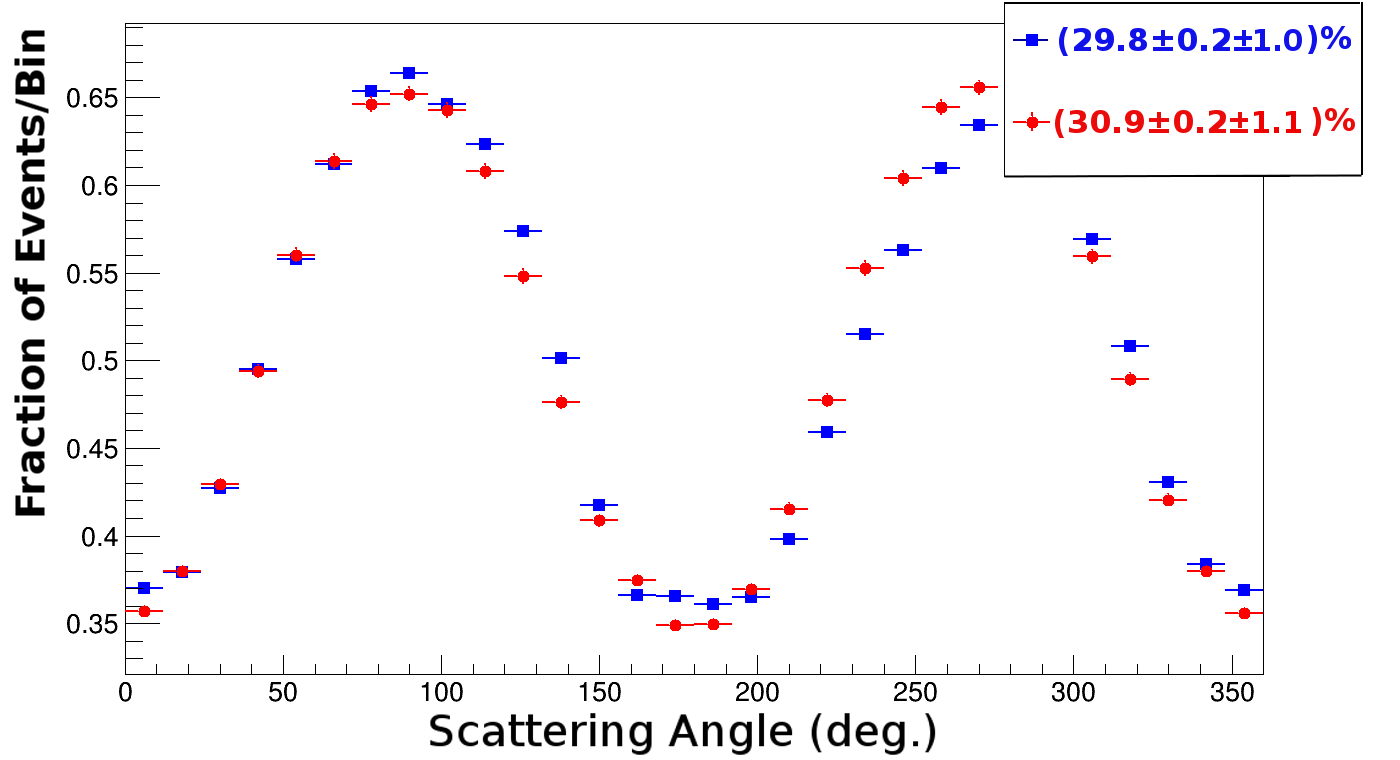}
  \caption{The measured polarized modulation curves for a 110 keV beam for 30 degrees off-axis measurement corrected by the simulated sample (squares) and the simulated polarized modulation curve corrected using the same simulated sample (circles). The data points only contain the statistical error. The fitted modulation factor is shown for each histogram. }
\label{30_off}
\end{figure}

\begin{figure}[h!]
  \centering
    \includegraphics[width=10 cm]{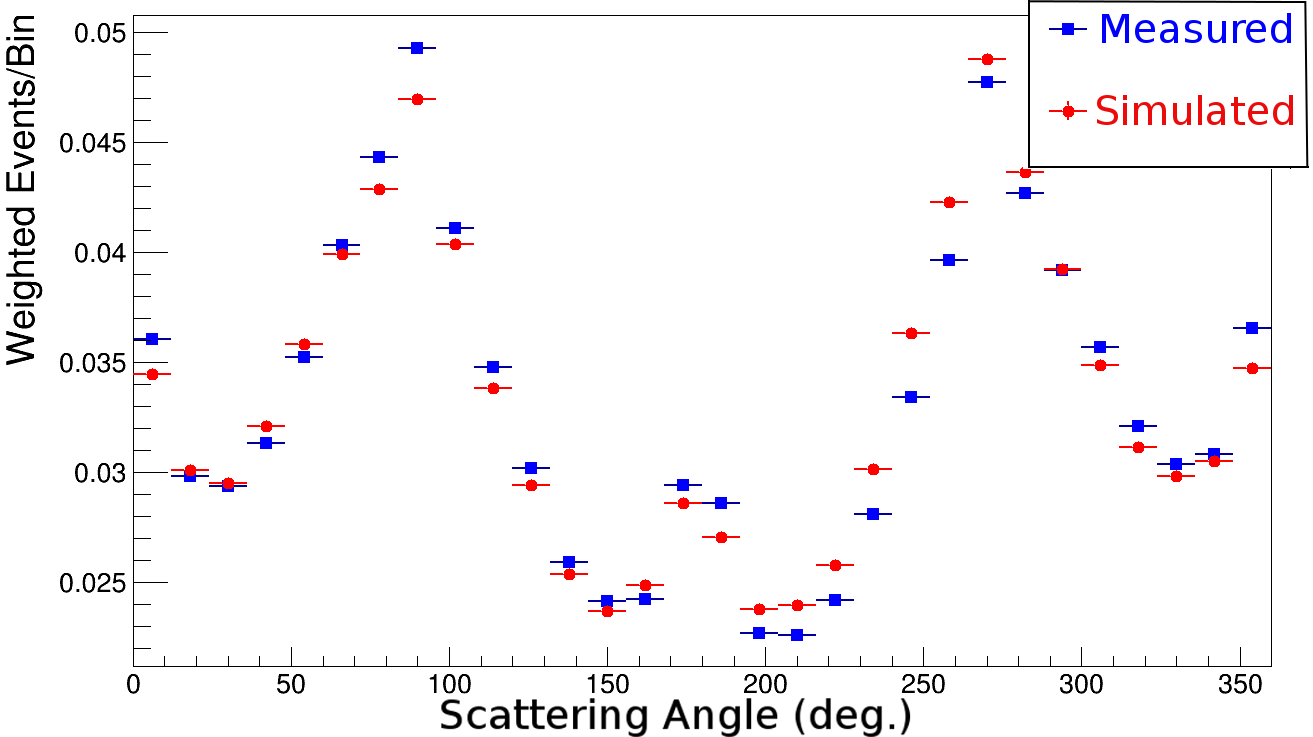}
  \caption{The measured polarized modulation curves for a 110 keV beam for $60^\circ$ off-axis scan of the full detector from measurement (squares) and from simulation (circles).  The data points only contain the statistical error.}
\label{110_60off}
\end{figure}

The off-axis response of POLAR was measured by rotating the instrument by $30^\circ$ and $60^\circ$ around the x-axis for each energy. The full instrument was scanned similar to the on-axis measurements with the only difference that the motor speed was lowered resulting in a higher amount of statistics. Using this method only one polarization direction can be measured. However as was shown with the on-axis data $\mu_{100}$ can be extracted by dividing the measured sample by the simulated unpolarized sample. The measured and simulated modulation curves for $30^\circ$  as measured for $110\,\mathrm{keV}$ are shown in figures \ref{30}. The figure shows a clear $360^\circ$ component induced by the off-axis beam. It can also be observed that again some instrumental effects are not modeled perfectly in the simulation. As in the off-axis measurements the majority of the events take place in 5 DMs, instead of the 25 in the on-axis measurements, such effects are more prominent in these measurements resulting in a larger systematic error on these components. The $180^\circ$ degree component in the simulated and measured modulation curves are, however, in good agreement. When dividing the measured modulation by the unpolarized simulated modulation curve the final resulting modulation is therefore very similar to that acquired when dividing the simulated polarized curve by a simulated unpolarized curve. The unpolarized and polarized simulated curves were produced from different samples. Despite still containing some geometrical effects the corrected measured curve shows good compatibility with the simulated curve as can be seen in figure \ref{30_off}. The resulting modulation is $(29.8\pm 0.2 \pm1.0)\%$ for the measured results and $(30.9\pm 0.2 \pm1.1)\%$ from the simulated results. The same procedure was followed for the $60^\circ$ off-axis measurement for which the uncorrected curves are shown in figure \ref{110_60off}. The value of $\mu_{100}$ obtained by dividing the measured sample by the simulated unpolarized sample is $(22.7\pm0.1\pm0.8)\%$ while doing the same exercise using the purely simulated sample the result was $(23.2\pm0.2\pm0.8)\%$. The results were therefore again found to be in good agreement.

\subsection{60 and 80 keV}

The measured and simulated modulation for the four different scans taken with the 110 keV beam are summarized in table \ref{table:110}. The $\mu_{100}$ values in the table referred to as simulated are the result of simulated polarized samples corrected by the simulated unpolarized samples. The measured values for $\mu_{100}$ for on-axis are from the measured polarized samples corrected by the measured unpolarized samples, whereas for off-axis these values were created by correcting the measured polarized samples by the simulated unpolarized samples. As the off-axis measured results are corrected with simulated data their systematic errors are larger than for the on-axis data which was corrected with measured data. For 60 keV the results are summarized in a similar way in table \ref{table:60}, while for 80 keV they are summarized in table \ref{table:80}.

It can be observed that the modulation factors obtained through simulations and measurements are in good agreement. The presented results show that using the less than optimum calibration parameters applied here the absolute uncertainty on the reconstructed value of $\mu_{100}$ remains on the order of $1\%$ for all the measurements. These results show that simulation data can therefore be used to remove the geometric and instrumental effects from in-orbit modulation curves where obtaining an unpolarized curve from measurements is not possible. 

\begin{table}
\begin{center}

\begin{tabular}{ |c|c|c|c| } 
 \hline
 Scan & Measurement ($\%$) & Simulated ($\%$) & Absolute Difference ($\%$)\\ 
 \hline
 On-axis (Vertical) & $34.0\pm0.1\pm0.1$ & $33.4\pm0.2\pm1.1$ & $0.6\pm0.3\pm1.2$ \\ 
 On-axis (Horizontal) & $34.4\pm0.1\pm0.1$ & $34.9\pm0.2\pm1.1$  & $0.5\pm0.3\pm1.2$ \\ 
 Off-axis ($30^\circ)$ & $29.8\pm0.2\pm1.0$ & $30.9\pm0.1\pm1.1$  & $1.1\pm0.3\pm2.1$ \\ 
 Off-axis ($60^\circ)$ & $22.7\pm0.1\pm0.8$ & $23.2\pm0.2\pm0.8$  & $0.5\pm0.3\pm1.6$ \\ 
 \hline
\end{tabular}
\caption{Summary of the modulation factors acquired using a 110 keV beam. Both the measured and simulated modulation factors are shown together with the difference between these two. For all values the fitting error is given first followed by the systematic error.}
\label{table:110}
\end{center}

\end{table}

\begin{table}
\begin{center}

\begin{tabular}{ |c|c|c|c| } 
 \hline
 Scan & Measurement ($\%$) & Simulated ($\%$) & Absolute Difference ($\%$)\\ 
  \hline
 On-axis (Vertical) & $21.7\pm0.1\pm0.1$ & $22.3\pm0.5\pm0.8$ & $0.6\pm0.6\pm0.9$\\ 
 On-axis (Horizontal) & $22.6\pm0.1\pm0.1$ & $22.9\pm0.5\pm0.8$ & $0.3\pm0.6\pm0.9$\\ 
 Off-axis ($30^\circ)$ & $18.8\pm0.5\pm0.6$ & $19.5\pm0.3\pm0.7$ & $0.7\pm0.8\pm1.2$\\ 
 Off-axis ($60^\circ)$ & $14.6\pm0.3\pm0.5$ & $14.6\pm0.5\pm0.5$ & $0.0\pm0.8\pm1.0$ \\ 
 \hline
\end{tabular}
\caption{Summary of the modulation factors acquired using a 60 keV beam. Both the measured and simulated modulation factors are shown together with the difference between these two. For all values the fitting error is given first followed by the systematic error.}
\label{table:60}
\end{center}

\end{table}

\begin{table}
\begin{center}

\begin{tabular}{ |c|c|c|c| } 
 \hline
 Scan & Measurement ($\%$) & Simulated ($\%$) & Absolute Difference ($\%$)\\ 
 \hline
 On-axis (Vertical) & $30.2\pm0.1\pm0.1$ & $30.8\pm0.2\pm1.1$ & $0.6\pm0.3\pm1.2$ \\ 
 On-axis (Horizontal) & $29.8\pm0.1\pm0.1$ & $29.1\pm0.1\pm1.0$ & $0.7\pm0.2\pm1.1$\\ 
 Off-axis ($30^\circ)$ & $25.9\pm0.1\pm0.7$ & $25.3\pm0.3\pm0.8$ & $0.6\pm0.4\pm1.5$\\ 
 Off-axis ($60^\circ)$ & $18.3\pm0.3\pm0.5$ & $19.4\pm0.3\pm0.7$ & $1.1\pm0.6\pm1.2$\\ 
 \hline
\end{tabular}
\caption{Summary of the modulation factors acquired using a 80 keV beam. Both the measured and simulated modulation factors are shown together with the difference between these two.}
\label{table:80}
\end{center}

\end{table}

\section{Conclusions}

POLAR is a new dedicated polarimeter launched in September 2016 with the goal of measuring the polarization parameters of hard X-rays from GRBs. Since these measurements are based on the non-uniformities of the trigger patterns in the instrument they require a detailed understanding of the instrument response. For this purpose POLAR has undergone calibration measurements at ESRF using different beam energies and different incoming photon angles. Furthermore a detailed simulation package was created. It was found that the simulations reproduce the instrumental and geometrical effects found in POLAR well within the systematic error which is dominated by the sub-optimum energy calibration opportunities. It was found that for the three different presented energies as well as for different incoming angles the simulated unpolarized data samples can be used to obtain modulation factors compatible with the measured values to within the errors which have a magnitude of approximately $1\%$. Furthermore it should be noted that the energy calibration in-orbit is expected to result in more precise gain calibrations which will result in more optimum simulations of geometrical effects reducing the systematic errors. 

\section{Acknowledgments}
We would like to acknowledge the personnel of the ID11 beam line at ESRF for their support during the performed beam tests, Dr. Maxime Chauvin for detailed discussions on which the used  digitization logic is based. The construction of POLAR was supported by the Swiss National Science Foundation, and by the Swiss Space Office through the ESA PRODEX program. 
This work is furthermore supported by National Science Center, Poland from grants 2015/17/N/ST9/03556 and from National Natural Science Foundation of China under the grant No. 11403026 and 11503028 and by the Strategic Priority Research Program of the Chinese Academy of Sciences, Grant No. XDB23040400.

\appendix
\section{\\}
	In order to apply the true crosstalk and gain these values have to be extracted from the measured gain and crosstalk values. 
	
	If $q_i$ is the output ADC level and $\Delta E_i$ is the visible deposited energy of each bar, $l$ is light yield, $c$ is photon collection efficiency, $g$ is the total gain of the PMT and other electronics and $m_{i,j}$ is the crosstalk matrix at the photon level, then we can write:
	
	\begin{equation}\label{equ:equ1}
	\begin{array}{lll}
	\begin{pmatrix}
	q_1 \\ q_2 \\ \vdots \\ q_{64}
	\end{pmatrix}
	& = &
	\begin{pmatrix}
	g_1    & 0      & \cdots & 0      \\
	0      & g_2    & \cdots & 0      \\
	\vdots & \vdots & \ddots & \vdots \\
	0      & 0      & \cdots & g_{64} \\
	\end{pmatrix}
	\begin{pmatrix}
	m_{1,1}  & m_{1,2}  & \cdots & m_{1,64}  \\
	m_{2,1}  & m_{2,2}  & \cdots & m_{2,64}  \\
	\vdots   & \vdots   & \ddots & \vdots    \\
	m_{64,1} & m_{64,2} & \cdots & m_{64,64} \\
	\end{pmatrix} \\ \\
	& &
	\begin{pmatrix}
	l_1c_1 & 0      & \cdots & 0            \\
	0      & l_2c_2 & \cdots & 0            \\
	\vdots & \vdots & \ddots & \vdots       \\
	0      & 0      & \cdots & l_{64}c_{64} \\
	\end{pmatrix}
	\begin{pmatrix}
	\Delta E_1 \\ \Delta E_2 \\ \vdots \\ \Delta E_{64}
	\end{pmatrix} \\ \\
	& = &
	\begin{pmatrix}
	g'_{1,1} m_{1,1}   & g'_{1,2} m_{1,2}   & \cdots & g'_{1,64}   m_{1,64}  \\
	g'_{2,1} m_{2,1}   & g'_{2,2} m_{2,2}   & \cdots & g'_{2,64}   m_{2,64}  \\
	\vdots             & \vdots             & \ddots & \vdots               \\
	g'_{64,1} m_{64,1} & g'_{64,2} m_{64,2} & \cdots & g'_{64,64} m_{64,64} \\
	\end{pmatrix}
	\begin{pmatrix}
	\Delta E_1 \\ \Delta E_2 \\ \vdots \\ \Delta E_{64}
	\end{pmatrix}
	\end{array}
	\end{equation} \\
	in above formula
	\begin{equation}
	g'_{i,j} = g_i l_j c_j \qquad i,j = 1,2,\cdots,64
	\end{equation}
	\begin{equation}
	\sum_{i=1}^{64}m_{i,j} = 1 \qquad j = 1,2,\cdots,64
	\end{equation}
	Here we assume the crosstalk of electronics to be negligible and assume that the non-uniformity mainly comes from the gain $g$.

	We can now write:
	\begin{equation}
	\small
	\begin{array}{lll}
	\begin{pmatrix}
	q_1 \\ q_2 \\ q_3 \\\vdots \\ q_{64}
	\end{pmatrix}
	& = &
	\begin{pmatrix}
	g'_{1,1} m_{1,1}   & g'_{1,2} m_{1,2}   & g'_{1,3} m_{1,3}   & \cdots & g'_{1,64}   m_{1,64} \\
	g'_{2,1} m_{2,1}   & g'_{2,2} m_{2,2}   & g'_{2,3} m_{2,3}   & \cdots & g'_{2,64}   m_{2,64} \\
	g'_{3,1} m_{3,1}   & g'_{3,1} m_{3,2}   & g'_{3,3} m_{3,3}   & \cdots & g'_{3,54}   m_{3,64} \\
	\vdots             & \vdots             & \vdots             & \ddots & \vdots               \\
	g'_{64,1} m_{64,1} & g'_{64,2} m_{64,2} & g'_{64,3} m_{64_3} & \cdots & g'_{64,64} m_{64,64} \\
	\end{pmatrix}
	\begin{pmatrix}
	\Delta E_1 \\ \Delta E_2 \\ \Delta E_3 \\ \vdots \\ \Delta E_{64}
	\end{pmatrix} \\ \\
	& = &
	\begin{pmatrix}
	1 & \frac{g'_{1,2} m_{1,2}}{g'_{2,2} m_{2,2}} & \frac{g'_{1,3} m_{1,3}}{g'_{3,3} m_{3,3}} & \cdots & \frac{g'_{1,64} m_{1,64}}{g'_{64,64} m_{64,64}} \\
	\frac{g'_{2,1} m_{2,1}}{g'_{1,1} m_{1,1}} & 1 & \frac{g'_{2,3} m_{2,3}}{g'_{3,3} m_{3,3}} & \cdots & \frac{g'_{2,64} m_{2,64}}{g'_{64,64} m_{64,64}} \\
	\frac{g'_{3,1} m_{3,1}}{g'_{1,1} m_{1,1}} & \frac{g'_{3,2} m_{3,2}}{g'_{2,2} m_{2,2}} & 1 & \cdots & \frac{g'_{3,64} m_{3,64}}{g'_{64,64} m_{64,64}} \\
	\vdots             & \vdots             & \vdots             & \ddots & \vdots               \\
	\frac{g'_{64,1} m_{64,1}}{g'_{1,1} m_{1,1}} & \frac{g'_{64,2} m_{64,2}}{g'_{2,2} m_{2,2}} & \frac{g'_{64,3} m_{64_3}}{g'_{3,3} m_{3,3}} & \cdots & 1 \\
	\end{pmatrix}
	\begin{pmatrix}
	g'_{1,1} m_{1,1}\Delta E_1 \\ g'_{2,2} m_{2,2}\Delta E_2 \\ g'_{3,3} m_{3,3}\Delta E_3 \\ \vdots \\ g'_{64,64} m_{64,64}\Delta E_{64}
	\end{pmatrix}
	\end{array}
	\end{equation}
	this can be defined by
	\begin{equation}
	\bm{Q} = \bm{X} \cdot \bm{Y} \cdot \bm{\Delta} \bm{E}
	\end{equation}
	\begin{equation}
	x_{i,j} = \frac{g'_{i,j} m_{i,j}}{g'_{j,j} m_{j,j}} \qquad i,j = 1,2,\cdots,64
	\end{equation}
	\begin{equation}
	y_j = g'_{j,j} m_{j,j} \qquad j = 1,2,\cdots,64
	\end{equation}
	\textbf{X} and \textbf{Y} are measured from the data. For simulations we however need $g'_{i,j}$ and $m_{i,j}$.
	\begin{equation}
	\sum_{i=1}^{64}m_{i,j} = 1 \quad\Rightarrow\quad
	m_{j,j}\sum_{i=1}^{64}\frac{m_{i,j}}{m_{j,j}} = g'_{j,j}m_{j,j}\sum_{i=1}^{64}\frac{x_{i,j}}{g'_{i,j}} = 1 \quad\Rightarrow\quad
	\sum_{i=1}^{64}\frac{x_{i,j}}{g'_{i,j}}= \frac{1}{y_j}
	\end{equation}
	\begin{equation}
	g'_{i,j} = g_il_jc_j \quad\Rightarrow\quad \sum_{i=1}^{64}\frac{x_{i,j}}{g_i}= \frac{l_jc_j}{y_j}
	\end{equation}
	\begin{equation}
	\begin{pmatrix}
	1        & x_{2,1}  & x_{3,1}  & \cdots & x_{64,1} \\
	x_{1,2}  & 1        & x_{3,2}  & \cdots & x_{64,2} \\
	x_{1,3}  & x_{2,3}  & 1        & \cdots & x_{64,3} \\
	\vdots   & \vdots   & \vdots   & \ddots & \vdots   \\
	x_{1,64} & x_{2,64} & x_{3,64} & \cdots & 1        \\
	\end{pmatrix}
	\begin{pmatrix}
	1/g_1 \\ 1/g_2 \\ 1/g_3 \\ \vdots \\ 1/g_{64}
	\end{pmatrix}
	=
	\begin{pmatrix}
	l_1c_1/y_1 \\ l_2c_2/y_2 \\ l_3c_3/y_3 \\ \vdots \\ l_{64}c_{64}/y_{64}
	\end{pmatrix}
	\end{equation}\\
	This equation can be solved. After getting $g_i$ using this, we can get
	\begin{equation}
	m_{i,j} = \frac{x_{i,j} g'_{j,j} m_{j,j}}{g'_{i,j}} = x_{i,j} \frac{y_j}{g_il_jc_j}
	\end{equation}
	Using this method, $g'_{i,j}$, the real gain, and $m_{i,j}$, the real crosstalk, can be separated.

\end{document}